\documentclass[aps,superscriptaddress, showpacs,preprintnumbers, superscriptaddress, nofootinbibt,twocolumn]{revtex4}
\usepackage{eurosym}
\usepackage{amsfonts}
\usepackage{amssymb,amsmath,enumitem}
\usepackage{graphicx,color,epstopdf}
\usepackage{subfigure}

\def\be{\begin{equation}}
\def\ee{\end{equation}}
\def\bea{\begin{eqnarray}}
\def\eea{\end{eqnarray}}

\newcommand{\f}[2]{\frac{#1}{#2}}

\begin{document}

\title{Distinguishing Brans-Dicke-Kerr type naked singularities and black holes with their thin disk electromagnetic radiation properties}
\author{Shahab Shahidi}
\email{s.shahidi@du.ac.ir}
\affiliation{School of Physics, Damghan University, Damghan,
	41167-36716, Iran,}
\author{Tiberiu Harko}
\email{tiberiu.harko@astro.ro}
\affiliation{Astronomical Observatory, 19 Ciresilor Street, 400487 Cluj-Napoca, Romania,}
\affiliation{Faculty of Physics, Babes-Bolyai University, Kogalniceanu Street,
Cluj-Napoca 400084, Romania,}
\affiliation{School of Physics, Sun Yat-Sen University, Xingang Road, Guangzhou 510275,
P. R. China,}
\author{Zolt\'{a}n Kov\'{a}cs}
\affiliation{Max-Fiedler-Str. 7, 45128 Essen, Germany }
\email{ kovacsz2013@yahoo.com}

\date{\today }

\begin{abstract}
The possible existence of naked singularities, hypothetical astrophysical
objects, characterized by a gravitational singularity without an event
horizon is still an open problem in present day astrophysics. From an observational point of view distinguishing
between astrophysical black holes and naked singularities also represents a
major challenge. One possible way of
differentiating naked singularities from black holes is through the
comparative study of thin accretion disks properties around these different
types of compact objects. In the present paper we continue the comparative
investigation of accretion disk properties around axially-symmetric rotating
geometries in Brans-Dicke theory in the presence of a massless scalar field. The solution of the field equations contains the
Kerr metric as a particular case, and, depending on the numerical values of the model parameter $\gamma$, has also solutions corresponding to non-trivial black holes and naked singularities, respectively.
Due to the differences in the exterior geometries between black holes and
Brans-Dicke-Kerr naked singularities, the thermodynamic and electromagnetic
properties of the disks (energy flux, temperature distribution and
equilibrium radiation spectrum) are different for these two classes of
compact objects, consequently giving clear observational signatures that
could discriminate between black holes and naked singularities.
\end{abstract}

\pacs{03.75.Kk, 11.27.+d, 98.80.Cq, 04.20.-q, 04.25.D-, 95.35.+d}
\maketitle
\tableofcontents

\section{Introduction}

The full understanding of the nature and possible structure of massive objects with mass functions greater than 3-4$M_{\odot}$ is still an open problem for present day theoretical astrophysics. The standard assumption about such objects is that they must be black holes, that is, objects whose surface is covered by an event horizon. Black holes result from the collapse of the stellar matter, when the gravitational effects cannot be counterbalanced by the baryonic pressure \cite{Ruff}. However, this scenario may not be the only alternative to the gravitational collapse. For example, quark stars in the Color-Flavor-Locked (CFL) phase can have masses in the range of $3.8M_{\odot}$ and $6M_{\odot}$, respectively, and thus they may be possible stellar mass black hole candidates \cite{K1}.
On the other hand it may be possible that during the gravitational collapse the vacuum breaks down, leading to the formation of gravastars, hypothetic objects that can  be described by the Schwarzschild metric, but without a
Schwarzschild horizon, while their inside region consists of a de Sitter type core \cite{G1,G2,G3, G4}.  Bosons stars \cite{Bose1} could also represent an alternative for the standard black hole picture.  The detection of the gravitational wave events \cite{Abbott1,Abbott2,Abbott3} strongly points towards black hole - black hole merger events, that could lead  to the measurability of the properties of binary black holes using gravitational waves. For recent discussions on the present situation in black hole physics, and of the possible alternatives to black holes see \cite{Pani1} and \cite{Pani2}, and references therein. It was also argued that giving an observational proof for the existence of a black-hole horizon by using electromagnetic waves is essentially impossible \cite{La}.

From a theoretical point of view the investigation of the final fate of matter, after the gravitational collapse of an initially
regular distribution of matter, represents one of the most important problems in general relativity. The first fundamental question would be to determine  under what kind of initial conditions the gravitational collapse ends in the formation of a black hole. However, it turns out that the final state
of the gravitational collapse is not necessarily always a black hole, and, depending on the initial conditions, a naked singularity can also
form as the end state of the collapse \cite{N1,N2,N3,N4,N5}. For reviews of the naked singularity problem see \cite{R1} and \cite{R2}. Hence one must also answer to the question if physically realistic collapse solutions of the Einstein gravitational equations that lead to the formation of naked singularities do correspond to some natural objects, observable by astronomical methods.  If found, such compact astrophysical bodies would be counterexamples of the cosmic censorship hypothesis, proposed by Roger Penrose \cite{Pe69}, and which conjectures that in asymptotically flat space-time event horizons always cover curvature singularities.

We can formulate the cosmic censorship conjecture either in a strong sense (in a physically appropriate geometry naked singularities cannot exist), or in a weak sense (even if such singularities do exist they are securely covered by an event horizon, and hence they cannot communicate with far-away observers). Since Penrose' s proposal, there have been many attempts to prove the conjecture (see \cite
{Jo93} and references therein for the early works in this field). But so far no proof of the conjecture has been presented. Still the analysis of the cosmic censorship conjecture is a very active field of research \cite{CS1,CS2,CS3,CS4,CS5,CS6,CS7,CS8,CS9,CS10,CS11,CS12,CS13,CS14,CS15,CS16,CS17,CS18,CS19,CS20,CS21,CS22,CS23,CS24,CS25,CS26,CS27,CS28,CS29,CS30,CS31,CS32,CS33,CS34}.

In \cite{CS1} it was suggested that the advanced Laser Interferometer Gravitational-wave Observatory would be able to detect violations of the cosmic censorship conjecture and of the no-hair theorem, since they limit the spin-to-mass-squared ratio of a Kerr black hole, and for    a non-rotating black hole suggests a particular value for the tidal Love number.  The behavior of massless scalar fields in the exterior of Reissner-Nordstr\"{o}m-de Sitter black holes was studied in \cite{CS16}. Their decay rates are governed by quasinormal modes of the black hole, and a detailed description of the linear scalar perturbations of the black holes was given.  Moreover, it was conjectured that the Strong Cosmic Censorship is violated in the near extremal regimes. In \cite{CS17} the suggestion that cosmic censorship in four-dimensional Einstein-Maxwell-$\Lambda$
theory would be removed if charged particles (with sufficient charge) were present was investigated.   The strong cosmic censorship hypothesis may be violated by nearly extremal Reissner-Nordstr\"{o}m-de Sitter black holes, since perturbations of such a black hole decay sufficiently rapidly so that the perturbed spacetime can be extended across the Cauchy horizon as a weak solution of the equations of motion.

The question of whether the introduction of a charged scalar field can save the strong cosmic censorship, which is violated by near-extremal Reissner-Nordstr\"{o}m-de Sitter black holes, was investigated in \cite{CS23}. Even so,  there is always a neighborhood of extremality in which strong cosmic censorship is violated by perturbations arising from smooth initial data. Counterexamples to cosmic censorship were discussed in \cite{CS24}.  The nonlinear Einstein-Maxwell-scalar field equations with a positive cosmological constant, under spherical symmetry, were solved numerically in \cite{CS26}, and it was found that mass inflation does not occur in the near extremal regime, indicating that nonlinear effects cannot save the Strong Cosmic Censorship Conjecture. For other recent investigations of the weak and strong cosmic censorship conjecture see \cite{CS27, CS28,CS29,CS30,CS31,CS32,CS33,CS34}.

The stability of the naked singularities in General Relativity has also been intensively investigated. In \cite{r1} it was shown that the negative mass Schwarzschild spacetime, which has a naked singularity, is perturbatively unstable. This result was obtained  by introducing a modification of the  Regge - Wheeler - Zerilli approach to black hole perturbations, and by showing the existence of exact exponentially growing solutions to the linearized Einstein's equations. Super-extremal black hole space-times (either with charge larger than mass or angular momentum larger than mass), which contain naked singularities, are unstable under linearized perturbations \cite{r2}. The evolution of the gravitational perturbations in a non globally hyperbolic background was considered in \cite{r3}, leading to the completion of the proof of the linear instability of the Schwarzschild naked singularity. This result was also supported by the numerical solutions of the linearized gravitational field equations. The exterior static region of a Reissner-Nordstr\"{o}m black hole is stable \cite{r4}. On the other hand the interior static region is unstable under linear gravitational perturbations \cite{r4}, with the field perturbations generically exciting a mode that grows exponentially in time. This result provides support to the strong cosmic censorship conjecture \cite{r4}.

The possible existence of unstable axisymmetric modes in Kerr space times was investigated in \cite{r5} by showing the existence of  exponentially growing solutions of the Teukolsky equation. Thus it follows that the stationary region beyond a Kerr black hole inner horizon is unstable under gravitational linear perturbations, and a Kerr space-time with angular momentum larger than its square mass, which has a naked singularity, is unstable.
 The gravitational-wave emission from the quasi-circular, extreme mass ratio inspiral of compact objects of mass $m_0$ into massive objects of mass $M>>m_0$ whose external metric is identical to the Schwarzschild metric, except for the absence of an event horizon, was studied in \cite{r6}, under the assumption  that such an object is a nonrotating thin-shell gravastar. For small values of the gravastar compactness the radiated power  carries the signature of the microscopic properties of the physical surface that replaces the event horizon. In \cite{r7} it was shown that both the interior region of a Kerr black hole $r<M-\sqrt{M^2-a^2}$ and the $a^2 > M^2$ Kerr naked singularity admit unstable solutions of the Teukolsky equation for any value of the spin weight.  The existence of the unstable modes is related to the so-called time machine region, where the axial Killing vector field is timelike, and the Teukolsky equation changes its character from hyperbolic to elliptic.

Hence, presently, despite the large number of studies in the field, the validity of the cosmic censorship conjectures are still a matter of debate, with many examples and counterexamples trying to provide support to its validity, or to unsubstantiate it. One important direction of research would be to try to confirm/infirm its soundness by using observational methods. In this context the possible detection of a naked singularity would give the final proof of the invalidity of the cosmic censorship conjecture. Such a possibility may be offered by the study of accretion phenomena.

Most astrophysical objects growth by mass accretion. The almost universal presence of interstellar matter generally leads to the formation around compact objects of accretion disks.  The emission of the radiation from the disk is determined by the external gravitational potentials of the central massive object, which in turn are essentially determined by its nature - neutron star, quark star, black hole, or naked singularity, for example. Hence the astrophysical observations of the emission spectra from accretion disks may lead to the possibility of directly testing the physical and astrophysical properties of the compact general relativistic objects that have generated the disk via their gravitational field. Modified gravity theories, like for example, $f(R)$ gravity, brane world models, or Horava-Lifshitz theory can be constrained and tested, using thin accretion disk properties \cite{mg1,mg2,mg3,mg4,mg5}. Wormhole geometries indicate significant differences in their disk accretion emission properties \cite{mg6,mg7, mg8}.  Gravastars can also be differentiated from ordinary black holes by using their accretion disk properties \cite{mg8}, while the electromagnetic properties of accretion disks around static (non-rotating) and  rotating neutron, quark, fermion and boson stars have been analyzed in \cite{mg9,mg10,mg11, mg11a, mg12,mg13,mg14,mg15,mg16,mg17,mg18,mg19}.

The possibility that naked singularities may be observationally distinguishable from their black hole counterparts by using the properties of the electromagnetic emissions of their thin disks was first proposed in \cite{KoHa10}. The specific astrophysical and astronomical signatures of the naked singularities have attracted have been extensively investigated in the literature.  In \cite{ns1} it was shown that a slowly evolving gravitationally collapsing perfect fluid cloud can asymptotically reach  a static spherically symmetric equilibrium configuration with a naked singularity at the center. The disk around the naked singularity is much more luminous than the one around the corresponding black hole, with the disk around the naked singularity having a spectrum with a high frequency power law segment that carries a major fraction of the total luminosity. Ultra-high-energy collisions of particles falling freely from rest at infinity can occur in the field of near-extreme Kehagias-Sfetsos naked singularities, with the efficiency of the escaping created ultrarelativistic particles and the energy efficiency of the collisional process relative to distant observers significantly lowered due to the large gravitational redshift \cite{ns2}. The lensing properties of the supermassive Galactic center of the Milky Way Galaxy, described as a naked singularity, were considered in \cite{ns3}. The observational properties of the Kehagias-Sfetsos naked singularities were further investigated in \cite{ns4}, \cite{ns5}, and \cite{ns6}, respectively. Tidal forces in naked singularity and black hole backgrounds were considered in \cite{ns7}, and the Roche limits were computed.  The redshift and properties of the shadow depend crucially on whether the final outcome of the complete gravitational collapse is a black hole or a naked singularity \cite{ns8}. Photons traveling from past to future null infinity through a collapsing object could provide an observational signature capable of differentiating between the formation of a globally naked singularity and the formation of an event horizon \cite{ns9}. The efficiency of the Keplerian accretion disks for all braneworld Kerr-Newman spacetimes was determined in \cite{ns10}. The precession of the spin of a test gyroscope due to the frame dragging by the central spinning body may be an important test for the existence of Kerr naked singularities. For Kerr black hole, the precession frequency becomes arbitrarily high, blowing up as the event horizon is approached, while in the case of a naked singularity, this frequency remains always finite and well behaved \cite{ns11,ns12}.  The periastron precession for a spinning test particle moving in nearly circular orbits around naked singularities was investigated in \cite{ns13}.

To distinguish a rotating Kiselev black hole from a naked singularity the critical values of the quintessential and spin parameters were studied in \cite{ns14}. Using the spin precessions one can  differentiate black holes from naked singularities. The possibility of discriminating black holes and naked singularities with iron line spectroscopy was investigated, for the case of the Janis-Newman-Winicour metric, in \cite{ns15}.  The iron line shapes in the reflection spectrum of a disk around a Janis-Newman-Winicour singularity were compared with the iron line shapes expected in the spectrum of a Kerr black hole. It turns out that Janis-Newman-Winicour singularities cannot mimic fast-rotating Kerr black holes, observed at a low or moderate inclination angle. The properties of spherical photon orbits in the field of Kerr naked singularities confined to constant Boyer-Lindquist radii were studied in \cite{ns16}.  The possibility of distinguishing rotating naked singularities from Kerr-like wormholes by their deflection angles of massive particles was investigated in \cite{ns17}. The comparison of the shadows cast by Schwarzschild black holes with those produced by two classes of naked singularities that result from gravitational collapse of spherically symmetric matter was performed in \cite{ns18}.  The possibility of differentiating  a Kerr-like black hole and a naked singularity in perfect fluid dark matter via precession frequencies was considered in \cite{ns19}. Circular orbits in Kerr-Taub-NUT spacetime and their implications for accreting black holes and naked singularities were analyzed in \cite{ns20}.  The optical appearance and the apparent radiation flux of a geometrically thin and optically thick accretion disk around the static Janis-Newman-Winicour naked singularity was studied in \cite{ns21}.  It was found that for the Janis-Newman-Winicour solution the accretion disk appears smaller, while its emission is characterized by a higher peak of the radiation flux.  Images of thin accretion disks around black holes and two classes of naked singularity spacetimes were comparatively studied in \cite{ns22}. The images obtained from naked singularity models significantly differ from those of black holes. The possibility that M87* might be a superspinar, that is, an object described by the Kerr solution and spinning so rapidly that it violates the Kerr bound by having $\left|a_*\right|>1$, was investigated in \cite{ns22a}. It was found that within certain regions of parameter space, the inferred circularity and size of the shadow of M87* do not exclude this  possibility.

A numerical algorithm for ray tracing in the external spacetimes of spinning compact objects characterized by arbitrary quadrupole moments was presented in \cite{ns23}. These objects correspond to non-Kerr vacuum solutions, and they can be used to test the no-hair theorem in conjunction with observations of accreting black holes. Allowing for the quadrupole moment of the spacetime to take arbitrary values leads to observable effects in the profiles of relativistic broadened fluorescent iron lines from geometrically thin accretion disks.
The effects induced by external magnetic fields on the observed thermal spectra and iron line profiles of thin accretion disks formed around Kerr black holes and naked singularities were considered in \cite{ns23}.  A numerical scheme able to calculate thermal spectra of magnetized Page-Thorne accretion disks formed around rotating black holes and naked singularities was developed, which can also be used to probe the cosmic censorship conjecture. Two different magnetic field configurations, uniform and dipolar, respectively, were considered. Observed synthetic line profiles of the 6.4 keV fluorescent iron line were also obtained.

 In \cite{ns25} it was shown that external magnetic fields produce  observable modifications on the thermal energy spectrum and the fluorescent iron line profile.  Comparison of the theoretical models with observational data can be used to probe the cosmic censorship conjecture. By using a ray-tracing algorithm to calculate the light curves and power spectra of hot spots on the disk one can prove that the emission from a hot spot orbiting near the innermost stable circular orbit of a naked singularity in a dipolar magnetic field can be significantly harder than the emission of the same hot spot in the absence of such a magnetic field.

As pointed out in \cite{Chau}, the (conformally related) Krori-Bhattacharjee spacetime, used in \cite{KoHa10} to study the accretion disk properties of naked singularities, is not a vacuum Brans-Dicke solution of the gravitational field equations (see also \cite{reply}). However,  a rotating solution that generalizes the Kerr metric for a minimally coupled scalar field in the framework of the Brans-Dicke theory does exist, and it was obtained in \cite{sol}. In the conformal frame this solution reduces to the Kerr metric for a specific value of the model parameter $\gamma$, while for other values it describes naked singularity and black hole geometries, respectively.

It is the goal of the present paper to investigate the electromagnetic emission properties of thin disks in the Kerr-Brans-Dicke geometry obtained in \cite{sol}. More exactly, we would like to consider some observational possibilities that may distinguish naked singularities from different types of black holes. One such observational possibility is offered by the study of the properties of the thin accretion disks that form around rotating compact general relativistic objects. In the present approach we restrict our analysis to the cases of naked singularities and black holes,
respectively. To achieve our objectives we consider a comparative study of the geometrical and physical properties of thin accretion disks around the rotating naked singularity, and rotating black holes, obtained as a solution of the field equations of the Brans-Dicke theory in \cite{sol}, in the presence of a scalar field. This solution contains as a particular case the Kerr metric of general relativity. We will analyze the basic physical parameters describing the thin accretion disks, including the electromagnetic energy flux, the temperature distribution on the surface of the disk, and the spectrum of the emitted equilibrium radiation.

Since the exterior geometries of the naked singularities and black holes are distinct, the corresponding differences do determine significant deviations in the thermodynamic and electromagnetic properties of the disks (energy flux, temperature distribution and equilibrium radiation spectrum)  for different classes of compact objects.  Thus the observations of the electromagnetic signals from accretion disks may provide some clear observational signatures that may allow to discriminate, at least in principle, black holes from naked singularities, and between different types of black holes. On the other hand we would like to point out that the possible detection of the naked singularities by using the electromagnetic properties of the  accretion disks is an {\it indirect} method, which must be considered together with {\it direct} methods of observation of the "surface" of the considered naked singularity/black hole/naked candidates. An alternative method to discriminate between  different types of compact objects is represented by their lensing properties.

The present paper is organized as follows. The rotating vacuum solution in the Brans-Dicke theory, and its geometry is presented in Section \ref{sect1}. In Section \ref{sect2} we present the main physical parameters (specific energy, the specific angular momentum, and angular velocity) describing the motion of massive test particles in stable circular orbits in arbitrary stationary and axisymmetric geometries.  We review the properties of standard thin accretion disks in Section \ref{sect3}. The observational properties of thin accretion disks formed around the Kerr-Brans-Dicke type compact objects (energy flux, temperature distribution, radiation spectrum and Eddington luminosity) are discussed in Section {\ref{sect4}.  We discuss and conclude our results in Section \ref{sect5}.

\section{The Kerr solution in the Brans-Dicke theory}\label{sect1}

The action of the Brans-Dicke theory, in which Newton's gravitational
constant is a variable function determined by a scalar field $\phi $ so that
$G=1/\phi $, is given by \cite{sol}
\begin{align}\label{act}
S=\frac{1}{16\pi }\int \Bigg[ \varphi R&-\frac{\omega }{\varphi }g^{\alpha
\beta }\nabla _{\alpha }\varphi \nabla _{\beta }\varphi\nonumber\\& -V(\varphi )+\mathcal{L}_{m}%
\Bigg] \sqrt{-g}d^{4}x,
\end{align}%
where $\omega $ is the dimensionless Brans-Dicke parameter, $V(\varphi )$ is
the scalar field potential, and $L_{m}$ is the matter action. By varying the
action with respect to the components of the metric tensor and of the scalar
field we obtain the Brans-Dicke field equations as
\bea
G_{\mu \nu }&=&\frac{8\pi }{\phi }T_{\mu \nu }+\frac{\omega }{\varphi ^{2}}%
\left[ \nabla _{\mu }\varphi \nabla _{\nu }\phi -\frac{1}{2}g_{\mu \nu
}\nabla _{\alpha }\varphi \nabla ^{\alpha }\varphi \right]\nonumber\\
&+&\frac{1}{\varphi
}\left[ \nabla _{\mu }\nabla _{\nu }\phi -g_{\mu \nu }\square \varphi \right]
,
\eea
\begin{align}
\Box\phi=\f{8\pi}{2\omega+3}T^{(m)},
\end{align}
where we have assumed that the scalar field potential vanishes, $T_{\mu \nu
}=-\left( 2/\sqrt{-g}\right) \delta \left( \sqrt{-g}\mathcal{L}_{m}\right) /\delta
g^{\mu \nu }$ is the matter energy-momentum tensor, and $T^{(m)}=T_{\mu}^{\mu}$, respectively.

A Kerr-like rotating
vacuum solution of the above field equations of the  Brans-Dicke theory was
obtained in \cite{sol}. By performing a conformal transformation of the
metric $g_{\mu \nu }\rightarrow \tilde{g}_{\mu \nu }=\Omega ^{2}g_{\mu \nu }$%
, with $\Omega =\sqrt{G\varphi }$, and by redefining the scalar field as
\be
\tilde{\varphi}=\sqrt{\frac{ 2\omega +3} {16\pi G}}\ln \left( \frac{\varphi}
{\varphi _{0}}\right) ,
\ee
where $\varphi _{0}$ is the present value of the
gravitational constant, it follows that in the conformal frame the field equations can be writtens as
\begin{align}
R_{\mu \nu }& =\frac{1}{2}\nabla _{\mu }\tilde{\varphi} \nabla _{\nu }\tilde{\varphi} ,
\notag  \label{fe} \\
\Box \tilde{\phi} & =0,
\end{align}%
with the solution \cite{sol}
\begin{widetext}
\begin{align}\label{ds2}
ds^2&=-fdt^2-\f{4Mar}{\Sigma}\sin^2\theta dtd\phi+\left(r^2+a^2+\f{2Ma^2r}{\Sigma}\sin^2\theta\right)\sin^2\theta d\phi^2+\left(\frac{\Delta\sin^2\theta}{M^2}\right)^{|1-\gamma|}\Sigma\left(\f{dr^2}{\Delta}+d\theta^2\right),
\end{align}
\end{widetext}
where $\gamma $ is a constant, and where we have defined
\begin{eqnarray}
f(r,\theta ) &=&1-\frac{2Mr}{\Sigma },\qquad \Sigma (r,\theta
)=r^{2}+a^{2}\cos ^{2}\theta ,  \notag \\
\Delta (r) &=&r^{2}+a^{2}-2Mr,
\end{eqnarray}%
In the solution (\ref{ds2}), the parameter $\gamma $ is related to the Brans-Dicke parameter $\omega $ as
\be
|1-\gamma| =\frac{4}{(2\omega +3)}.
\ee
Also, the scalar field $\tilde{\varphi} $ can be obtained as
\begin{equation} \label{scal}
\tilde{\varphi} =\sqrt{|1-\gamma| }\ln \left( \frac{\Delta \sin ^{2}\theta }{M^{2}}%
\right).
\end{equation}%
It is worth noting that the special case $\gamma =1$ corresponds to the
Kerr black hole. The Kretchmann scalar $R_{\mu \nu \rho\sigma}R^{\mu \nu \rho\sigma}$ of the metric \eqref{ds2} can be obtained as
\begin{align}
R_{\mu \nu \rho\sigma}R^{\mu \nu \rho\sigma}=\Sigma^{-6}\left(\frac{\Delta}{m}\right)^{-2-2|1-\gamma|}g(r,\theta).
\end{align}
One can see from the above relation that the Kretchmann scalar $R_{\mu \nu \rho\sigma}R^{\mu \nu \rho\sigma}$ diverges for $\Sigma =0$. The function $g(r,\theta)$ is a regular function with the property that for $\gamma=1$ it has a form $g(r,\theta)\propto\Delta^2$. As a result one can see that for $\Delta=0$, which corresponds to $r=r_{\pm }$, with
\begin{equation}
r_{\pm }=M(1\pm \sqrt{1-a_{\ast }^{2}}),  \label{rp}
\end{equation}%
the Kretchmann scalar diverges, and we have a curvature singularity at $r=r_\pm$. However, from the line element \eqref{ds2} one can see  that for $0<\gamma <2$ we have a horizon at $r=r_{\pm
}$.
Hence, for $0<\gamma <1$ and $1<\gamma <2$ the curvature singularity at $r=r_+$ will be covered by the horizon $r_{+}$.

As a result it follows that for $-\infty <\gamma \leq 0$ and $2<\gamma<\infty$, respectively, the metric \eqref{ds2} describes the spacetime geometry of a naked singularity, with a total mass $M$, and an
angular momentum $J=aM=a_{\ast }M^{2}$. Here $a_{\ast }=J/M^{2}$ is the
dimensionless spin parameter.

To summarize our analysis, in the case of the metric (\ref{ds2}) we have a naked singularity in the range  $-\infty <\gamma \leq 0$, and $2<\gamma<\infty$, respectively, and a non-trivial black hole in the range $0<\gamma <1$ and $1<\gamma <2$.

The surface of infinite redshift is determined by the condition $f=0$, which defines the ergo-sphere of the rotating geometry \eqref{ds2}
as
\begin{align}
r_{s,n}=M(1\pm\sqrt{1-a^2_* \cos^2\theta})\:.  \label{rs}
\end{align}

The frame dragging frequency $\omega$ of this rotating solution can be obtained as
\begin{equation} \label{omega}
\omega =\frac{2Mar}{\Sigma(r^2+a^2)+2Ma^2r\sin^2\theta}\;,
\end{equation}
which has the same form as the frame dragging frequency of the Kerr black hole.

\section{Motion of test particles in stable orbits around rotating compact objects}\label{sect2}

In the present Section we will briefly review the basic results concerning the motion of massive test particles in arbitrary axisymmetric geometries, and then we will apply the obtained results to the case of the metric (\ref{ds2}), giving the analogue of the Kerr metric in the Brans-Dicke theory.

\subsection{The general formalism}

An arbitrary axisymmetric geometry can be generally described by a line element of
the form
\begin{align}\label{ds2rcoappr}
ds^{2}=g_{tt}c^2dt^{2}+2g_{t\phi }cdtd\phi +g_{rr}dr^{2}+g_{\theta \theta
}d\theta ^{2}+g_{\phi \phi }d\phi ^{2}\;.
\end{align}%

In Eq.~(\ref{ds2rcoappr}), due to the adopted symmetry of the spacetime,  the metric components $g_{tt}$, $g_{t\phi }$, $g_{rr}$, $g_{\theta
\theta }$ and $g_{\phi \phi }$ depend only on  $r$ and $%
\theta$. One can easily see that for the motion in the above geometry we have two conserved
quantities, the specific energy at infinity $\tilde{E}$, and the $z$%
-component of the specific angular momentum at infinity $\tilde{L}$, respectively, which
can be obtained as \cite{KoHa10}
\begin{equation}\label{1}
g_{tt}\dot{t}+g_{t\phi}\dot{\phi}=-\widetilde{E},
\end{equation}
\begin{equation}  \label{2}
g_{t\phi}\dot{t}+g_{\phi \phi}\dot{\phi}=\widetilde{L},
\end{equation}
where a dot denotes derivative with respect to the affine parameter $\tau
$.

In the equatorial plane with $\theta=\pi/2$, one can obtain the geodesic
equations as
\begin{align}
\dot{t}&=\frac{\widetilde{E}g_{\phi \phi }+\widetilde{L}g_{t\phi }}{g_{t\phi
}^{2}-g_{tt}g_{\phi \phi }},  \notag \\
\dot{\phi}&=-\frac{\widetilde{E}g_{t\phi }+\widetilde{L}g_{tt}}{g_{t\phi
}^{2}-g_{tt}g_{\phi \phi }}.
\end{align}%
and
\begin{equation}  \label{geodeqs3}
g_{rr}\dot{r} ^{2}=V_{eff}\left(r,\widetilde{E},\widetilde{L}\right),
\end{equation}%
where we have defined \cite{KoHa10}
\begin{equation}
V_{eff}= -1+\frac{\widetilde{E}^{2}g_{\phi \phi }+2\widetilde{E}\widetilde{L}%
g_{t\phi }+\widetilde{L}^{2}g_{tt}}{g_{t\phi }^{2}-g_{tt}g_{\phi \phi }}\;.
\end{equation}

For circular orbits in the equatorial plane, we have $V(r)=0$ and $%
V_{,r}(r)=0$, which determine the specific energy $\widetilde{E}$, the
specific angular momentum $\widetilde{L}$ as a function of the angular
velocity $\Omega$ of particles as \cite{Shib}
\begin{eqnarray}
\widetilde{E} &=&-\frac{g_{tt}+g_{t\phi }\Omega }{\sqrt{-g_{tt}-2g_{t\phi
}\Omega -g_{\phi \phi }\Omega ^{2}}}\;,  \label{tildeE} \\
\widetilde{L} &=&\frac{g_{t\phi }+g_{\phi \phi }\Omega }{\sqrt{%
-g_{tt}-2g_{t\phi }\Omega -g_{\phi \phi }\Omega ^{2}}},  \label{tildeL} \\
\Omega &=&\frac{d\phi }{dt}=\frac{-g_{t\phi ,r}\pm\sqrt{(g_{t\phi
,r})^{2}-g_{tt,r}g_{\phi \phi ,r}}}{g_{\phi \phi ,r}},  \notag  \label{Omega}
\\
\end{eqnarray}
where in the definition of $\Omega$, the plus and minus signs correspond to
the direct and retrograded orbits, respectively.

Any stationary observer, moving along a world line $r=\mathrm{constant}$ and
$\theta=\mathrm{constant}$ with a uniform angular velocity $\Omega$, has a
four-velocity vector $u^{\mu}\propto(\partial/\partial
t)^{\mu}+\Omega(\partial/\partial{\phi})^{\mu}$, which lies inside the
surface of the future light cone. Therefore, we have the condition \cite{KoHa10}
\begin{equation}
g_{tt} + 2 \Omega g_{t\phi} + \Omega^2 g_{\phi\phi} \leq 0.  \label{cond}
\end{equation}
The above relation puts a constraint on the value of the angular velocity as
$\Omega_{min}<\Omega<\Omega_{max}$, where we have defined
\begin{eqnarray}
\Omega_{min}&=&\omega-\sqrt{\omega^2-\frac{g_{tt}}{g_{\phi\phi}}}\:,
\label{Omin} \\
\Omega_{max}&=&\omega+\sqrt{\omega^2-\frac{g_{tt}}{g_{\phi\phi}}}\:,
\label{Omax}
\end{eqnarray}
and $\omega=-g_{t%
\phi}/g_{\phi\phi}$ is the frame dragging frequency.

The limiting case constraint of Eq.~\eqref{cond},
\begin{equation}  \label{rph}
g_{tt} + 2 \Omega g_{t\phi} + \Omega^2 g_{\phi\phi} = 0,
\end{equation}
gives the innermost boundary of the circular orbits for particles, $r_{ph}$,
called photon orbit. Circular orbits with $\widetilde{E}<1$ are bounded. The limiting case $%
\widetilde{E}=1$ gives the radius $r_{mb}$ of the marginally bound circular
orbit, that is, the innermost orbits.

The marginally stable circular orbits $r_{ms}$ around the
central object can be determined from the condition \cite{KoHa10}
\bea \label{stable}
\hspace{-0.6cm}V_{,rr}|_{r=r_{ms}}&=&
\frac{1}{g^2_{t\phi}-g_{tt}g_{t\phi}}\Bigg[\widetilde{E}^{2}g_{\phi \phi ,rr}+2\widetilde{E}\widetilde{L}g_{t\phi ,rr}+\nonumber\\
\hspace{-0.6cm}&&\widetilde{L}^{2}g_{tt ,rr}-
(g_{t\phi }^{2}-g_{tt}g_{\phi \phi})_{,rr}\Bigg]\Bigg|_{r=r_{ms}}=0. \nonumber\\
\eea
For stable circular orbits the condition $V_{,rr}<0$ must be satisfied. From a physical point of
view we can interpret the marginally stable orbit as the innermost boundary
of the stable circular orbits.

\subsection{Circular motions in the equatorial plane of the Brans-Dicke-Kerr naked singularity}

Inserting the metric components in the definitions of the specific energy,
angular momentum and of the angular velocity as given by Eqs.~\eqref{tildeE}-\eqref{Omega} into Eqs.~\eqref{rph} and \eqref{stable}, respectively, we obtain a set of algebraic equations for $r_{ms}$, $r_{mb}$ and $r_{ph}$. One should note that because
only the $g_{rr}$ and $g_{\theta\theta}$ components of the metric \eqref{ds2}
are different from the metric of the Kerr black hole, it follows that for the Brans-Dicke-Kerr metric in the conformal frame  $r_{ms}$, $r_{mb}$, $%
r_{ph}$ and $r_\pm$ are the same as for the Kerr metric of standard general relativity. Fig.~\ref{fig1}
shows the behavior of these radii as a function of $a_\star$.

\begin{figure}[h]
 \centering
\includegraphics[scale=0.46]{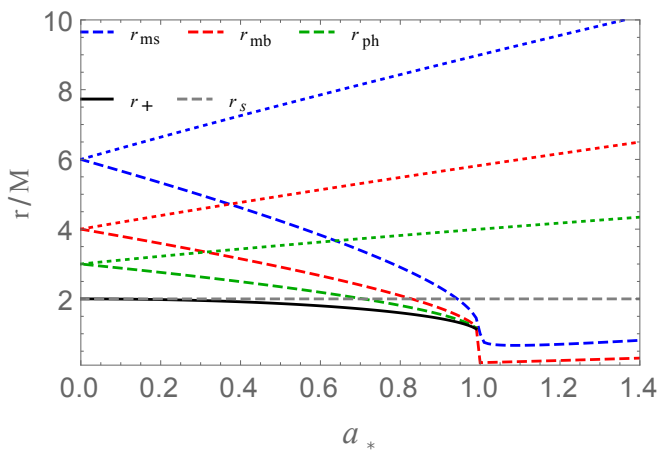}
\caption{Radii of the circular orbits around the compact general relativistic object with exterior metric given by Eq.~\eqref{ds2} as a
function of $a_{\star}$. The event horizon and the ergosphere are also
depicted. The dashed lines represents direct orbits and the dotted lines
represent retrograded orbits.}\label{fig1}
\end{figure}

Also, the values of the second derivative of the potential \eqref{stable} have
the same form as for the Kerr metric. In Fig.~\ref{fig2} we have plotted the
behavior of $V_{,rr}(r=r_{ms})$ for different values of $a_\star$. it is worth mentioning that for $a_\star>1$, the Kerr geometry describes a naked singularity. In this sense we will consider these cases in this paper to include this interesting case.
\begin{figure}[h]
 \centering
\includegraphics[scale=0.45]{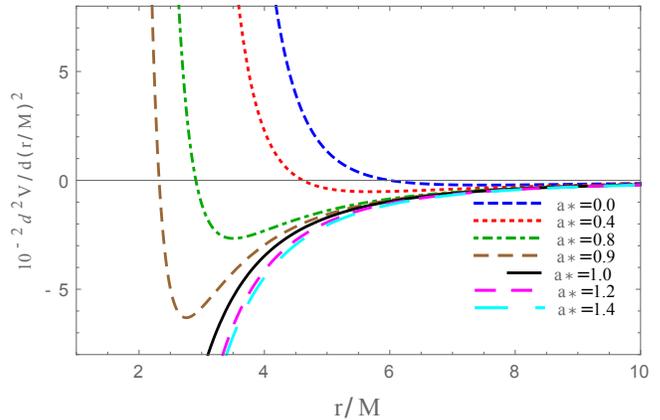}
\caption{Behavior of the second derivative of the potential of the Brans-Dicke-Kerr object given by Eq.~\eqref{stable}
with respect to $r/M$ for different values of $a_\star=0,0.4,0.7,0.9,1,1.2,1.4$.}\label{fig2}
\end{figure}

Hence all the results on the motion of test particles in the equatorial plane of the Kerr geometry \cite{Carter, Press, Shib} are also valid in the case of the considered Kerr-Brans-Dicke geometry. The expressions of the radii of the marginally stable orbits can be obtained analytically as \cite{Press,Shib}
\be
r_{ms}=\frac{GM}{c^2}\left[3+R_{ms}^{(2)}\pm\sqrt{\left(3-R_{ms}^{(1)}\right)\left(3+R_{ms}^{(1)}+2R_{ms}^{(2)}\right)}\right],
\ee
where $R_{ms}^{(1)}=1+\sqrt[3]{1-a_{\ast }^{2}}\left( \sqrt[3]{1+a_{\ast }}+\sqrt[3%
]{1-a_{\ast }}\right) $, and $R_{ms}^{(2)}=\sqrt{3a_{\ast }^{2}+\left(
R_{ms}^{(1)}\right) ^{2}}$, respectively. The positive sign corresponds to the retrograde orbits, while the negative sign describes the prograde (direct) motion. For $a_*=0$, from the above expression we reobtain the expression of the radius of the marginally stable orbits for the Schwarzschild metric, $r_{ms}=6GM/c^2$, while $a_*=1$  gives for the direct orbit $r_{ms}=GM/c^2$, while for the retrograde orbit in the Kerr geometry $r_{ms}=9GM/c^2$. From these considerations it seems that for $a_*=1$ the radius of the marginally stable orbit is located at the same radial coordinate as the horizon itself. However, one can show that in the Kerr geometry the radii of the marginally stable orbits are always greater than the horizon radius \cite{Press, Shib}.

If $a_*>1$ there are no singularities in the Kerr geometry, and we have the Kerr naked singularity spacetimes,  where no horizons do exist. In these types of Kerr geometries, the physical singularity is located at $r = 0$, and $\theta = \pi/2$, respectively. Hence the radii of the marginally stable orbits of the Kerr naked singularities can come closer to the central singularity, which would induce a significant effect on particle dynamics. The Kerr naked singularities have some specific properties that could differentiate them with respect to the Kerr black holes, like, for example, the properties of the spherical photon stable orbits confined to constant Boyer-Lindquist radius $r$ that could be pure prograde/retrograde, or with turning points in the azimuthal direction \cite{Stuch}.

\section{Electromagnetic effects in accretion disks gravitating around compact objects}\label{sect3}

In the following we will review the basics of the thin accretion disks theory in general relativity. Observationally, accretion disks are common objects, and they are observed as flattened astronomical structures, consisting of a rapidly rotating hot gas that slowly moves towards a central dense and massive object. The internal stresses and the dynamical friction of the disk matter generates heat, with a small fraction of it being converted into electromagnetic radiation that can escape from the disk surface, leading to the  cooling down of the disk. Therefore, once detected in the radio, optical or X-ray frequency bands, important information about the accretion disk physics can be obtained from the study of the electromagnetic spectrum of the disk radiation, and of its time variability. Hence important information about the physical processes in and near the disks can be obtained from observations.  In many cases the inner edge of the disk is positioned at the marginally stable orbit of the gravitational potential of the central object, with the hot gas having a Keplerian motion \cite{PaTh74, Th74}.

The general relativistic theory of mass accretion around compact objects was developed by Novikov and Thorne  in \cite{NoTh73}, by
extending the steady-state thin disk models introduced in
\cite{ShSu73}. In the Novikov-Thorne approach, which considered the case of the curved space-times, the equatorial
approximation was adopted for the stationary and axisymmetric geometry.
In the equatorial approximation it is assumed that the vertical size of the disk (defined along the $z$-axis)
is much smaller than its horizontal extension (defined along the
radial direction $r$). Equivalently, for a thin disk, the disk height $H$, equal to the maximum
half thickness of the disk, is assumed to be always much smaller than the characteristic radius $R$ of
the disk, $H \ll R$.

From a physical point of view the thin disk is assumed to be in hydrodynamical equilibrium, while the pressure gradient
and the vertical entropy gradient in the accreting matter are neglected. In the following we will adopt the steady state disk accretion model, in which
the mass accretion rate $\dot{M%
}_{0} $ is supposed to be constant in time. Moreover, all the physical quantities describing the properties
of the matter in the disk are averaged over a characteristic time scale $%
\Delta t$, and over the azimuthal angle $\Delta \phi =2\pi $.

With the use of the four dimensional conservation laws
of the rest mass, of the energy, and of the angular momentum of the disk matter, respectively,
we can obtain the structure equations of the thin disk.
The flux of the radiant energy released by the disk surface can be expressed as
\cite{PaTh74, Th74}
\begin{equation}
F(r)=-\frac{\dot{M}_0}{4\pi\sqrt{-g}} \frac{\Omega_{,r}}{(\widetilde{E}%
-\Omega\widetilde{L})^{2}} \int_{r_{ms}}^{r}(\widetilde{E}-\Omega\widetilde{L%
})\widetilde{L}_{,r}dr\;,  \label{F}
\end{equation}
where we have also assumed the no-torque inner boundary conditions \cite{PaTh74}, which implies that the torque vanishes at the inner edge of the disk, and where \begin{align}\label{det}
\sqrt{-g}=\left(\f{\Delta\sin^2\theta}{M^2}\right)^{1-\gamma}\Sigma(r,\theta)\sin\theta.
\end{align}

By supposing that in the steady-state thin disk the accreting matter is
in thermodynamical equilibrium, the radiation emitted by the surface of the disk can be approximated by a perfect black body radiation, described by the Planck distribution function $I(\nu )$. Hence the
energy flux can be obtained as $F(r)=\sigma _{SB}T^{4}(r)$, where $\sigma _{SB}$ is the
Stefan-Boltzmann constant, with the observed luminosity $L\left( \nu \right) $ having a redshifted black body spectrum, given by \cite{mg9}
\begin{equation}
L\left( \nu \right) =4\pi d^{2}I\left( \nu \right) =\frac{8\pi h \cos i}{ c^2 } \int_{r_{i}}^{r_{f}}\int_0^{2\pi}\frac{\nu^{3}_e r d\phi dr }{\exp \left( \nu_e/T\right) -1},\label{L}
\end{equation}
where $d$ is the distance to the source,   $i $ is the disk inclination angle (which in the following we will take it to be zero), while $r_{i}$ and $r_{f}$ denote the positions of the inner and outer edges of the disk,
respectively.

In our analysis of the disk properties around Kerr-Brans-Dicke compact objects we adopt the values $r_{i}=r_{ms}$ and $r_{f}\rightarrow \infty $, the last condition implying
that the flux generated by the disk surface vanishes at infinity.
 The frequency of the radiation emitted by the disk is given by $\nu_e=\nu(1+z)$, where the redshift factor $z$ can be obtained as  \cite{Lu79,BMT01}
\begin{equation}\label{31}
1+z=\frac{1+\Omega r \sin \phi \sin i }{\sqrt{ -g_{tt} - 2 \Omega g_{t\phi} - \Omega^2 g_{\phi\phi}}}\;.
\end{equation}
In the case of the Schwarzschild metric $ds^2=-(1-2M/r)dt^2+dr^2/(1-2M/r)+r^2\left(d\theta ^2+\sin ^2\theta d\phi ^2\right)$,  by taking into account that $\Omega =\left(M/r^3\right)^{1/2}$, the redshift factor is given by \cite{Lu79}
\be\label{32}
1+z=\frac{1+\left(M/r^3\right)^{1/2}b\sin \phi \sin i}{\sqrt{1-3M/r}},
\ee
where $b$ is the impact parameter. On the other hand the deflection angle of light at infinity $\phi _{\infty}$ by a massive object can be obtained as $\phi _{\infty}=2(P/Q)^{1/2}\left[K(k)-F\left(\zeta _{\infty},k\right)\right]$, where $P$ is the periastron distance, $Q^2=\left(P-2M\right)\left(P+6M\right)$, $k=\left(Q-P+3M/Q\right)^{1/2}$, $\sin ^2\zeta _{\infty}=\left(Q-P+2M\right)/\left(Q-P+6M\right)$, while $K(k)$ and $F\left(\zeta _{\infty},k\right)$ are the complete integral of modulus $k$ and and the eliptic integral of modulus $k$ and argument $\zeta _{\infty}$, respectively \cite{Lu79}. In the limit of $P\rightarrow 3M$ for the total deviation of a light ray $\mu =2\phi_{\infty}-\pi$ we obtain the approximate relation $b=5.19695M+3.4823Me^{-\mu}$ \cite{Lu79}. As one can see from Eq.~(\ref{32}), in the limit $r\rightarrow 3M$, the redshift factor takes very large values (tending to infinity), while the total deviation of the light has finite (and relatively small) values. Even for the minimum value $r_{min}=6M$ of the inner radius of the disk around a Schwarzschild black hole the redshift factor is much bigger than the deflection angle $\mu \approx 4M/b$.  The same qualitative results are also valid in the case of the Kerr geometry. Generally, the factor $-g_{tt}-2\Omega g_{t\phi}-\Omega ^2 g_{\phi \phi}$ becomes smaller when approaching the inner edge of the disk, or the event horizon, and this leads to a significant increase in the redshift factor, as compared to the bending of light.
Hence, by taking into account the above results, in the following we will neglect in Eq.~(\ref{L}) the effects of the gravitational light bending by the central massive object \cite{Lu79,BMT01}.

An important parameter characterizing accretion disks is the efficiency $\epsilon $, indicating the capability  of the central object to convert rest mass into the radiation emitted by the disk.  The parameter $\epsilon $
is defined as the ratio of the rate of the
 energy of the photons escaping from the disk surface to infinity, and the energy rate at which mass-energy is transported to the central object. If we assume that the entire emitted electromagnetic energy can travel to infinity, then $\epsilon $ is determined only by the specific energy estimated at the marginally stable orbit $r_{ms}$, so that $\epsilon = 1 - \left.\widetilde{E}\right|_{r=r_{ms}}$.

For a Schwarzschild black holes $\epsilon $ is of the order of 6\%, and this value is independent of the photon capture by the black hole. For rapidly
rotating black holes,  $\epsilon$ is around 42\%, while by taking into account photon capture the efficiency is 40\% for
the Kerr geometry.

\section{Observational signatures of Brans-Dicke-Kerr type geometries}\label{sect4}

In the following we will analyze the electromagnetic emission properties of the accretion disks around Brans-Dicke-Kerr compact objects, for which the exterior geometry is described by Eq.~(\ref{ds2}).

\subsection{Electromagnetic properties of the disk}

The emission and physical properties of accretion disks are mainly characterized by the energy flux, the temperature, and the disk luminosity. We will consider each of these properties for the accretion disks located in the gravitational field of the Brans-Dicke-Kerr type compact object.

\subsubsection{The energy flux profiles}

\begin{figure*}[tbp]
 \centering
\includegraphics[scale=0.47]{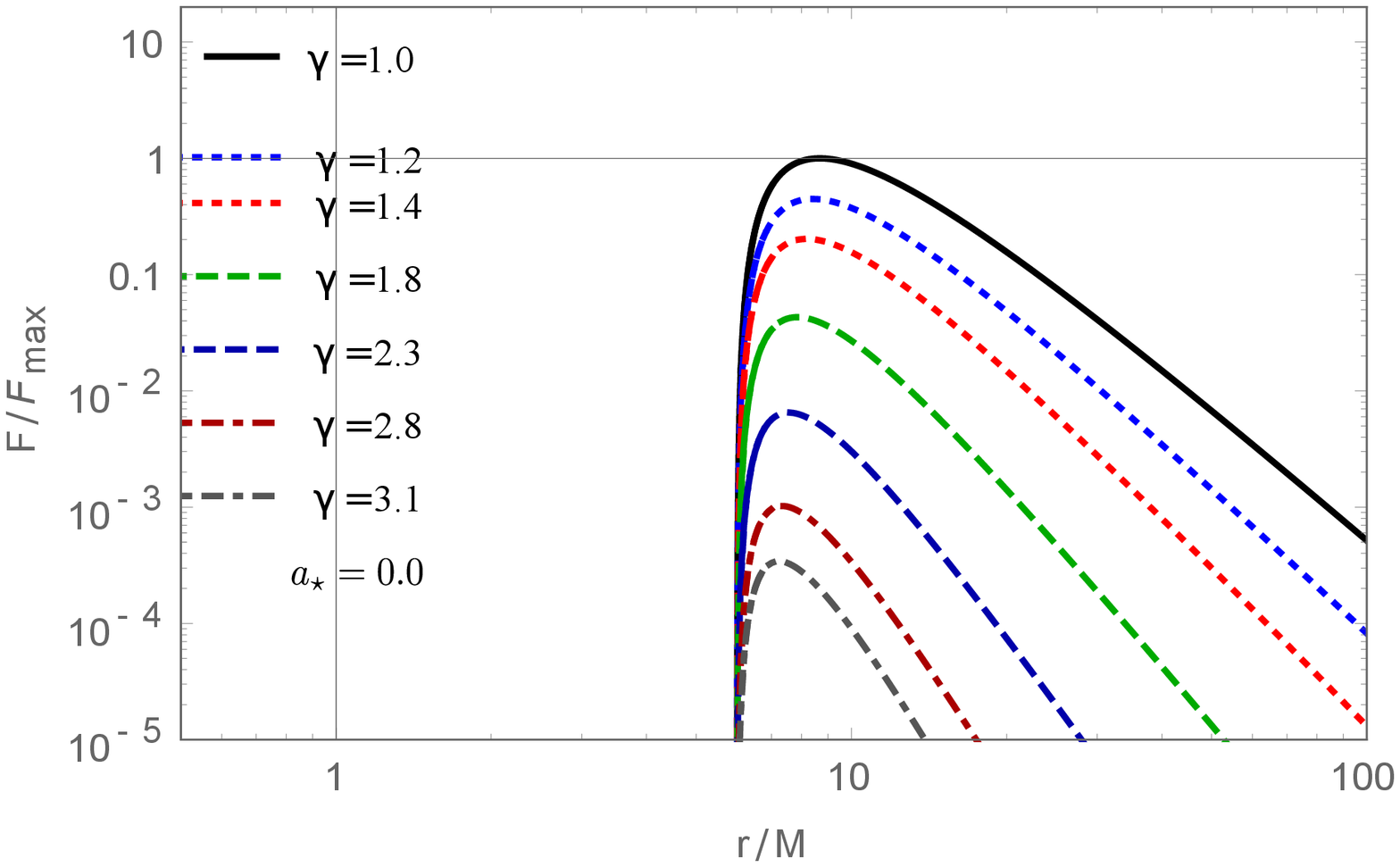}~\includegraphics[scale=0.47]{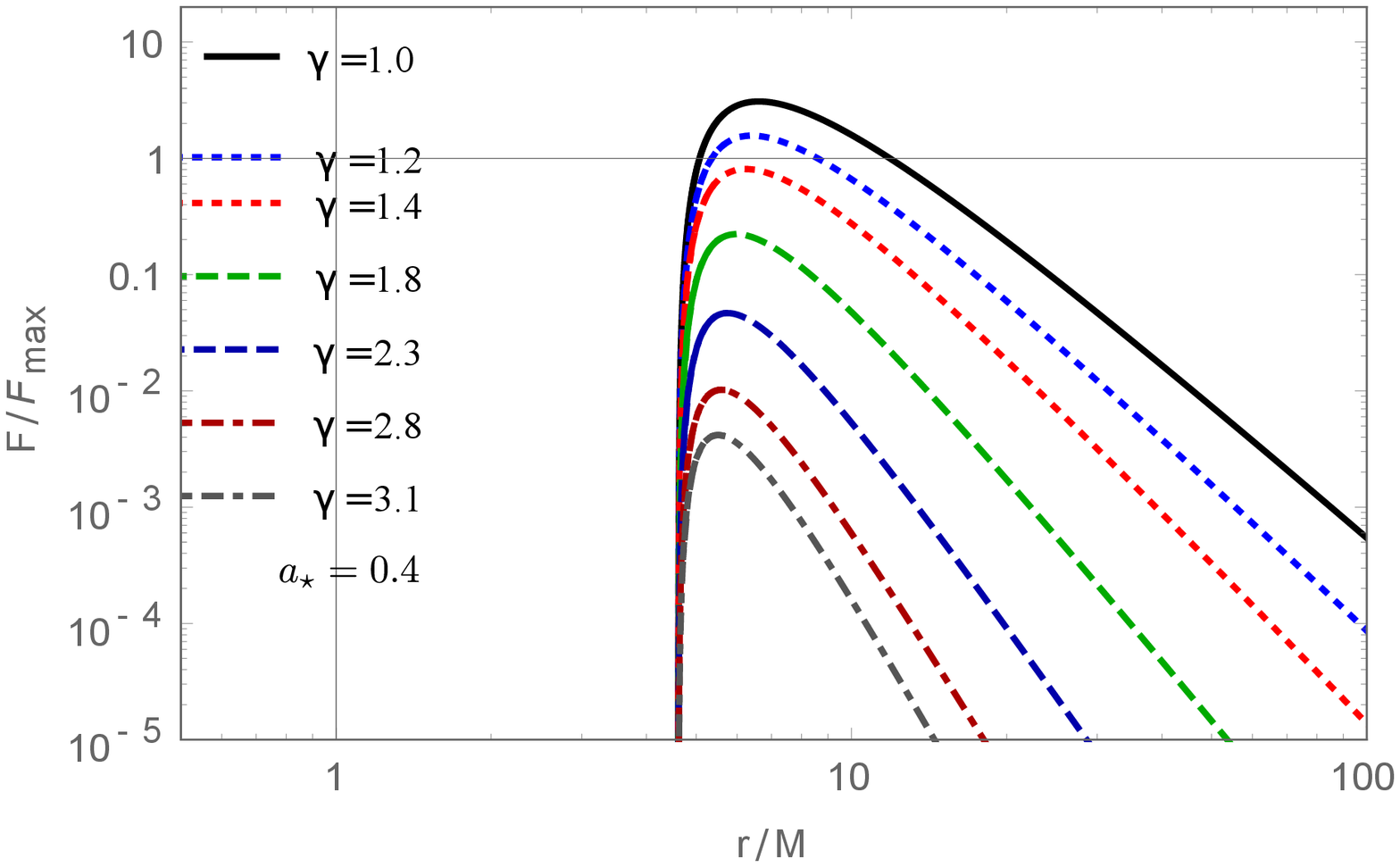}\\%
\noindent  \vspace{0.2cm}\newline
\includegraphics[scale=0.47]{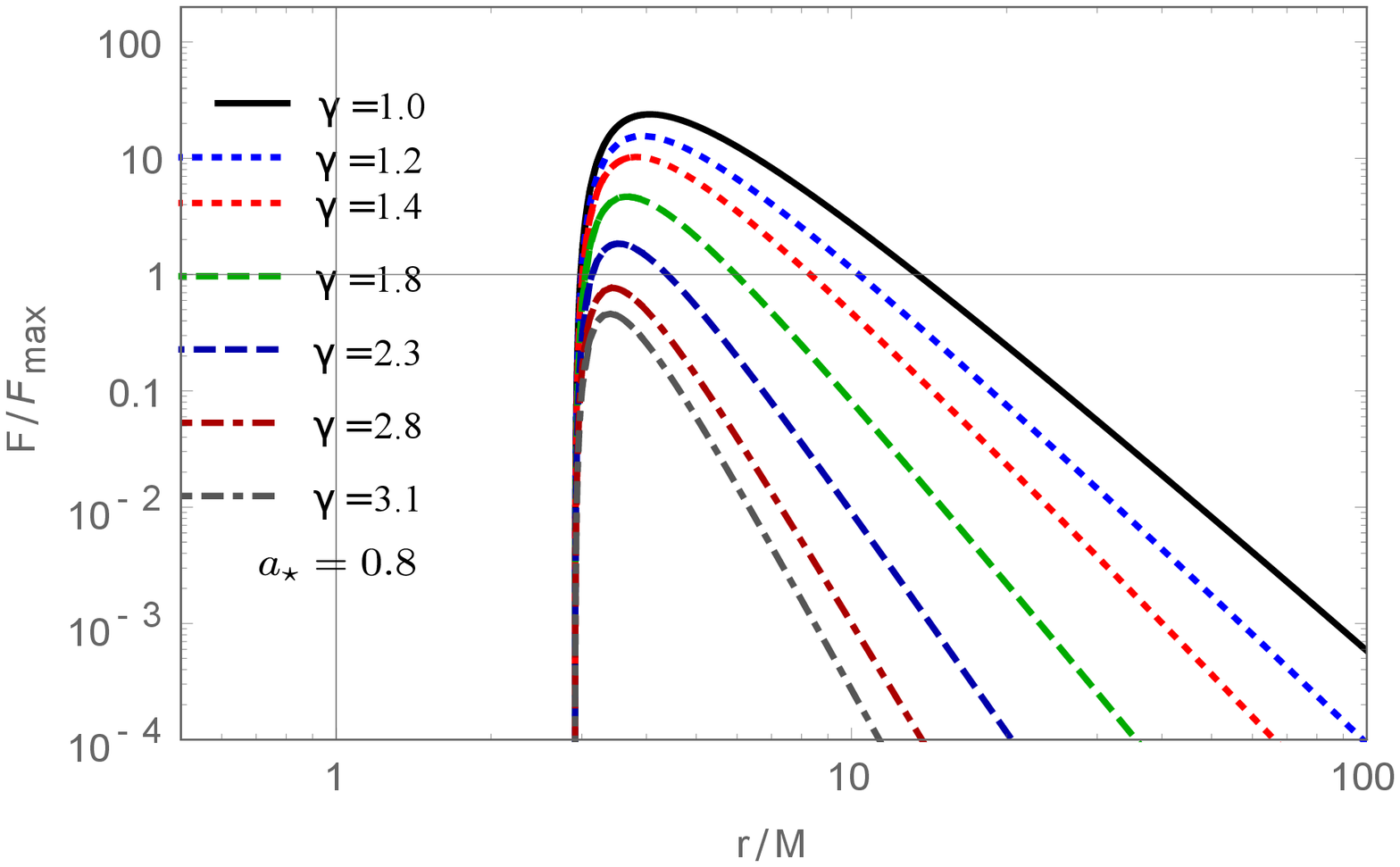}~\includegraphics[scale=0.47]{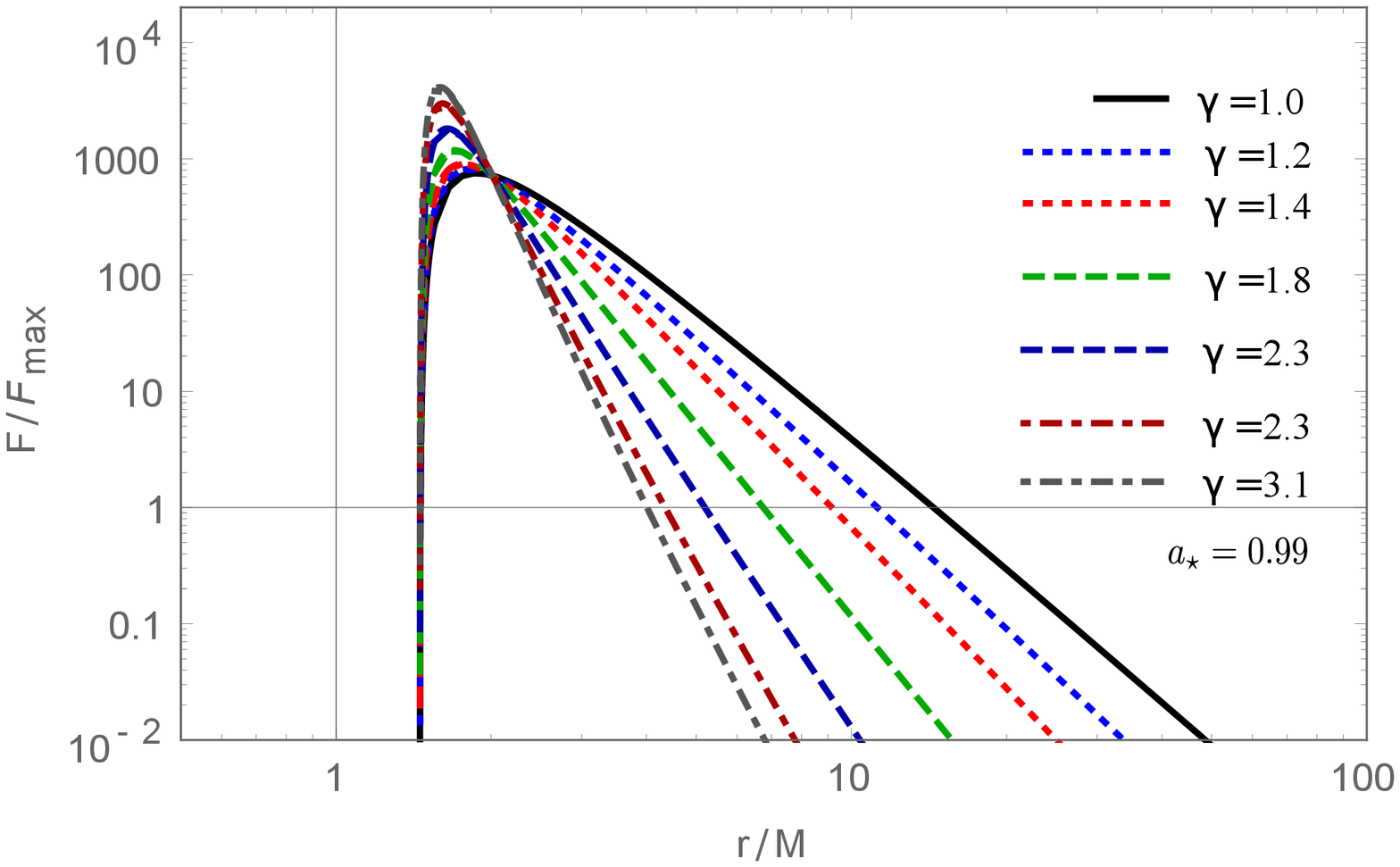}
\noindent  \vspace{0.2cm}\newline
\includegraphics[scale=0.47]{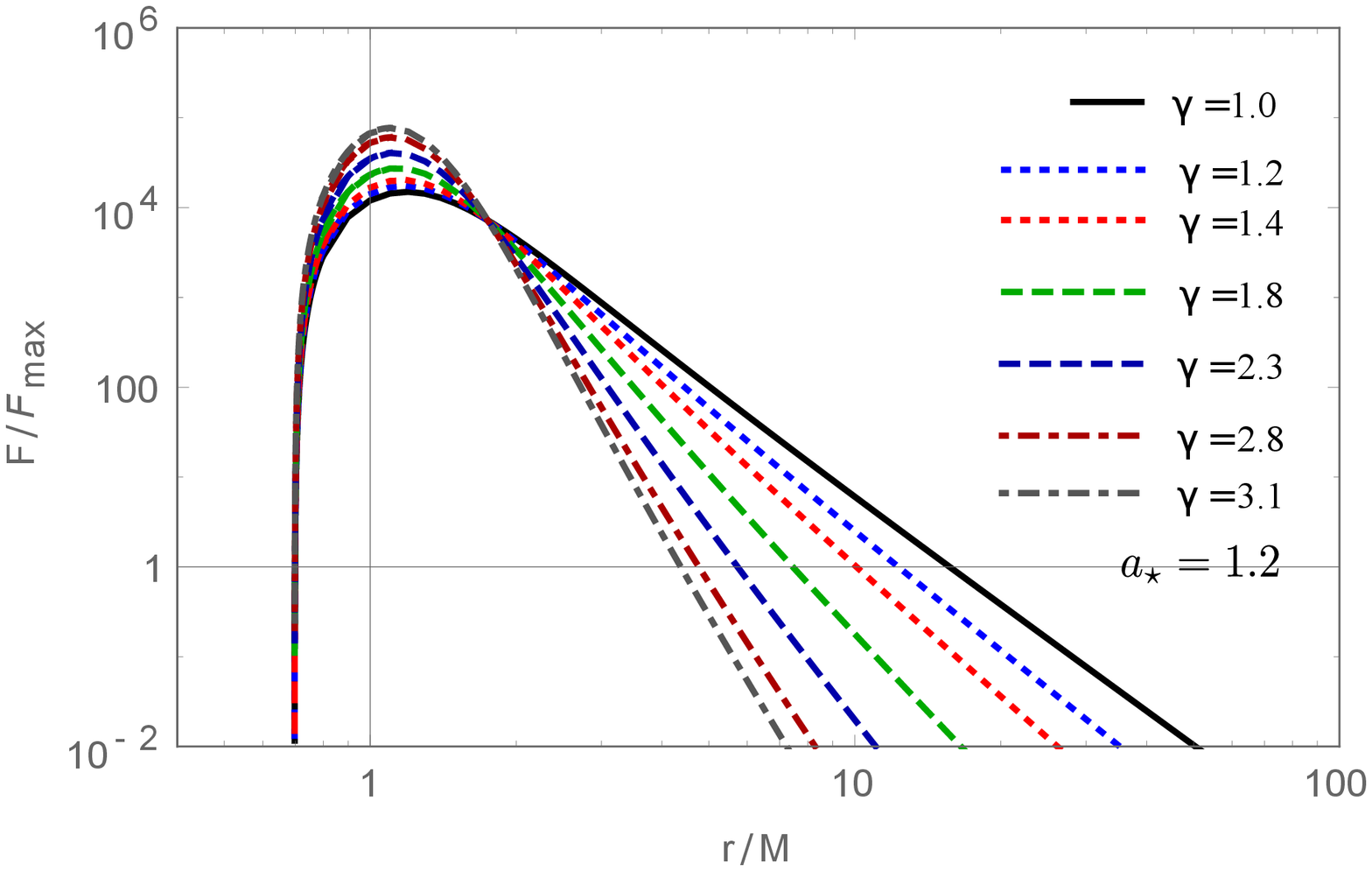}~\includegraphics[scale=0.47]{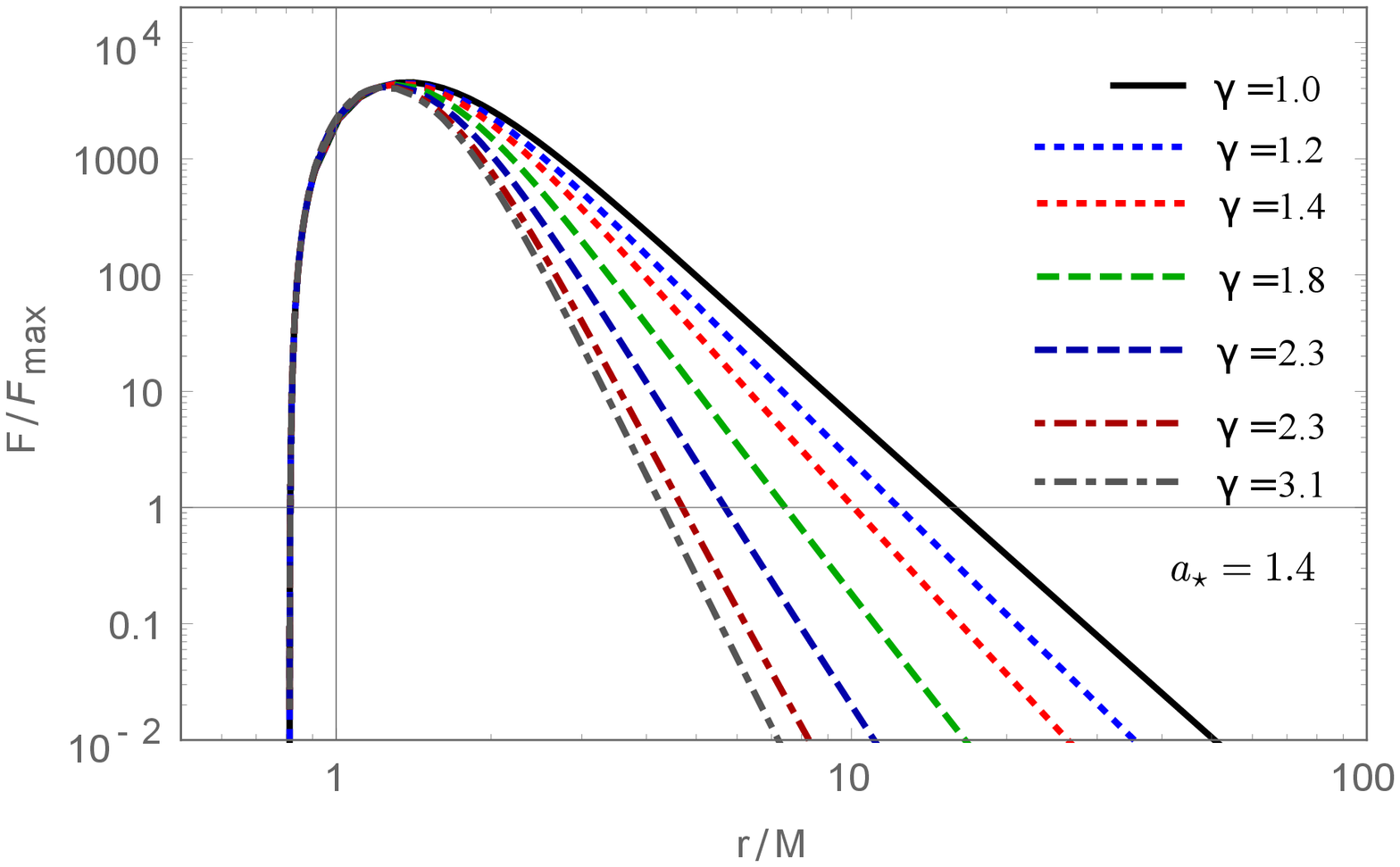}
\caption{The log-log plot of the normalized energy flux $F(r)/F_{max}$ with respect to $%
r/M$ for $a_\star=0,\;0.4,\;0.8,\;0.99,1.2,1.4$, and  for $\gamma=1$ (Kerr black hole for $a_\star\leq1$, and naked Kerr singularity for $a_\star>1$), $\gamma%
=1.2,\;1.4,\;1.8$ (non-trivial black holes) and $\gamma=2.3,\;2.8,\;3.1$ (naked singularities), respectively. Here $F_{max}$ is the flux of the Schwarzschild disk with $\gamma=1$, and $a_\star=0$, respectively.}\label{fig3}
\end{figure*}

In Figs.~\ref{fig3} we have plotted the normalized energy flux profiles, computed from Eq.~(\ref{F}), for
different values of parameter $\gamma$, and $a_\star$. Note that the expression for the
energy flux differs from the expression corresponding to the Kerr metric only through the expression of the metric
determinant, given by Eq.~\eqref{det}.  Moreover, the values of $r_{ms}$ for all three types of compact objects described by the metric (\ref{ds2}) are the same, and therefore in this geometry the matter cannot approach the naked singularity. Also we have normalized the energy flux by a factor
$F_{max}$, which is the maximum value of the disk flux for the Schwarzschild metric with $\gamma=1$ and $a_\star=0$.

As for the physical parameters of the configurations we have adopted the numerical values $a_{\star}=0$ (corresponding to the static case), $a_{\star}=0.4$, $a_{\star}=0.80$, $a_{\star}=0.99$, corresponding to the extreme rotation limit of the Kerr black hole, and $a_*=1.2$ and $a_*=1.4$, respectively, with the last two values describing for $\gamma =1$ the Kerr naked singularities. The values of $\gamma $ have been chosen in three distinct ranges, to describe three types of different astrophysical objects: the Kerr black hole and naked singularity, corresponding to $\gamma =1$, the non-trivial black hole, with an event horizon, obtained for $\gamma =1.2$, $\gamma =1.4$, and $\gamma =1.8$, respectively, and the Kerr-Brans-Dicke type naked singularity, which appears for $\gamma =2.3$, $\gamma =2.8$, and $\gamma =3.1$.

As a general result of our investigations we find that there is a significant difference in the energy fluxes from the disks rotating around these three types of compact objects. Interestingly enough, in the cases of the static black hole, and for $a_{\star}=0.4$ and $a_{\star}=0.8$, the maximum value of the flux is obtained for the standard Kerr black hole of general relativity. The thermal energy fluxes from the disks around the non-trivial Kerr-Brans-Dicke type black holes, and of the naked singularities are significantly smaller than the Kerr flux, the differences being of the order of three to four orders of magnitude in the case of the $\gamma =3.1$ naked singularity. However, with increasing spin, the maximal flux is also increasing, and tends to reach the maximal Kerr value. Another interesting phenomenon is that for higher values of the spin, the locations of the maxima of the energy fluxes shift toward lower radii, located closer to the inner edge of the disk. This effect is stronger for the naked singularities of the Kerr-Brans-Dicke type solution.

The behavior of the energy fluxes show a drastic change in the extreme rotation limit $a_{\star}=0.99$, presented in the middle right panel of Fig.~\ref{fig3}. In this case the energy fluxes  from the disks around naked singularities and non-trivial black holes can exceed with almost one order of magnitude the emission of the Kerr disk. Moreover, the shift of the maximum values towards lower radii indicates that most of the electromagnetic radiation comes from the inner edge of the disk.

The differences between the physical properties of the fluxes become even more important in the case of the comparison of the Kerr naked singularities, corresponding to $a_*>1$,  with the Brans-Dicke-Kerr type naked singularities and non-trivial black holes, depicted in the bottom panel of Fig.~\ref{fig3}. The maximum of the energy flux is shifted significantly towards the central singularity, and it is attained for values of $r/M$ of the order of 1, a value smaller than the one corresponding to the case of the extremely rotating Kerr black hole. Moreover, the maximum energy flux of the Kerr naked singularity is smaller than the values obtained for the Brans-Dicke-Kerr type naked singularities and non-trivial black holes. The maximum value of the flux increases with increasing $\gamma $, and for $a_*=1.2$ it exceeds by one order of magnitude the maximum value of the flux for the maximally rotating Kerr black hole with $a_*=0.99$. However, there is slight decrease in the maximum values of the fluxes with increasing $a_*$, a result due to the fact that the radii of the marginally stable orbits tend to decrease with increasing $a_*$ (for $a_*=1.20$, $r_{ms}=0.6983GM/c^2$, while $r_{ms}=0.8121GM/c^2$ for $a_*=1.40$). Another significant difference is related to the flux distribution over the disk. The flux decreases faster for the Brans-Dicke-Kerr singularities/black holes as compared to the Kerr naked singularity case, indicating that the main energy emission takes place in a limited area mostly concentrated in the inner region of the disk. This is in fact a general result valid for all the cases we have investigated.

The result that the flux maximum is higher for the rapidly rotating non-trivial black holes and naked singularities than for the Kerr black holes and naked singularities, even if it is integrated over a smaller surface area, is the direct consequence of the important differences in the metric determinants of the metrics, which, in the vicinity of the equatorial plane, characterizes the four-volume element in which the electromagnetic radiation flux is measured. For Kerr black holes in the equatorial approximation the expression $\sqrt{-g}=r^2$ holds, but from the expression (\ref{det}) of the determinant of the rotating Kerr-Brans-Dicke solution we obtain
\be
\sqrt{-g}=\left(\frac{\Delta}{M^2}\right)^{1-\gamma}r^2=\left(\frac{M^2}{r^2+a^2-2Mr}\right)^{\gamma -1}r^2, \gamma >1.
\ee
 Hence it follows that the function $\left(\Delta/M^2\right)^{1-\gamma}$ has a smaller value when we are approaching $r_{ms}$, and for large rotational velocities. Then it turns out that the four-volume element is much smaller for the non trivial black holes and for naked singularities as compared to the standard Kerr black hole case, and it gives much higher values in the energy flux integral (\ref{F}) for the former types of objects, even if the geometric properties determining $\Omega$, $\tilde{E}$ and $\tilde{L}$ are similar in the two cases.

\subsubsection{Temperature distribution}

\begin{figure*}[tbp]
\centering
\includegraphics[scale=0.47]{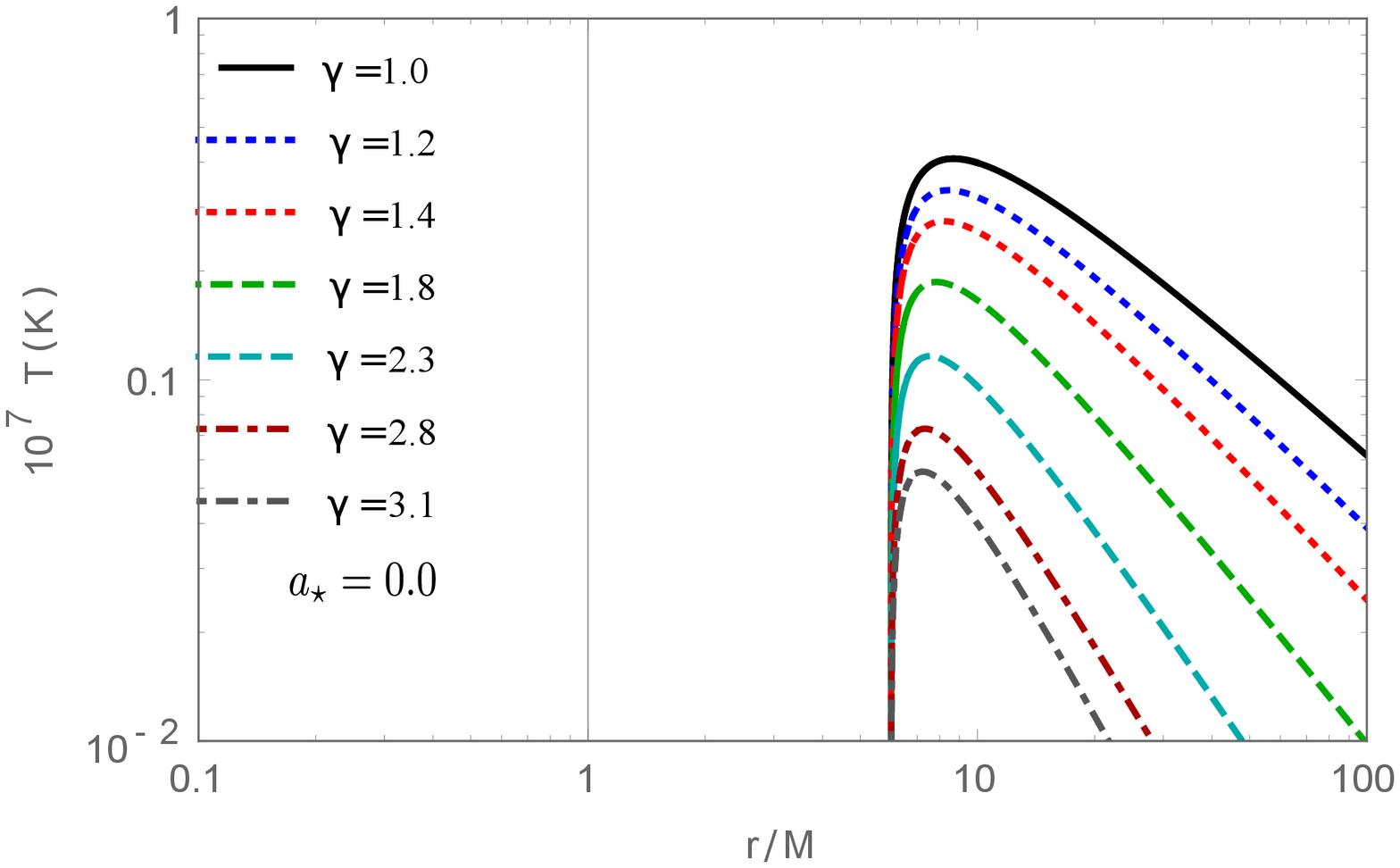}~\includegraphics[scale=0.47]{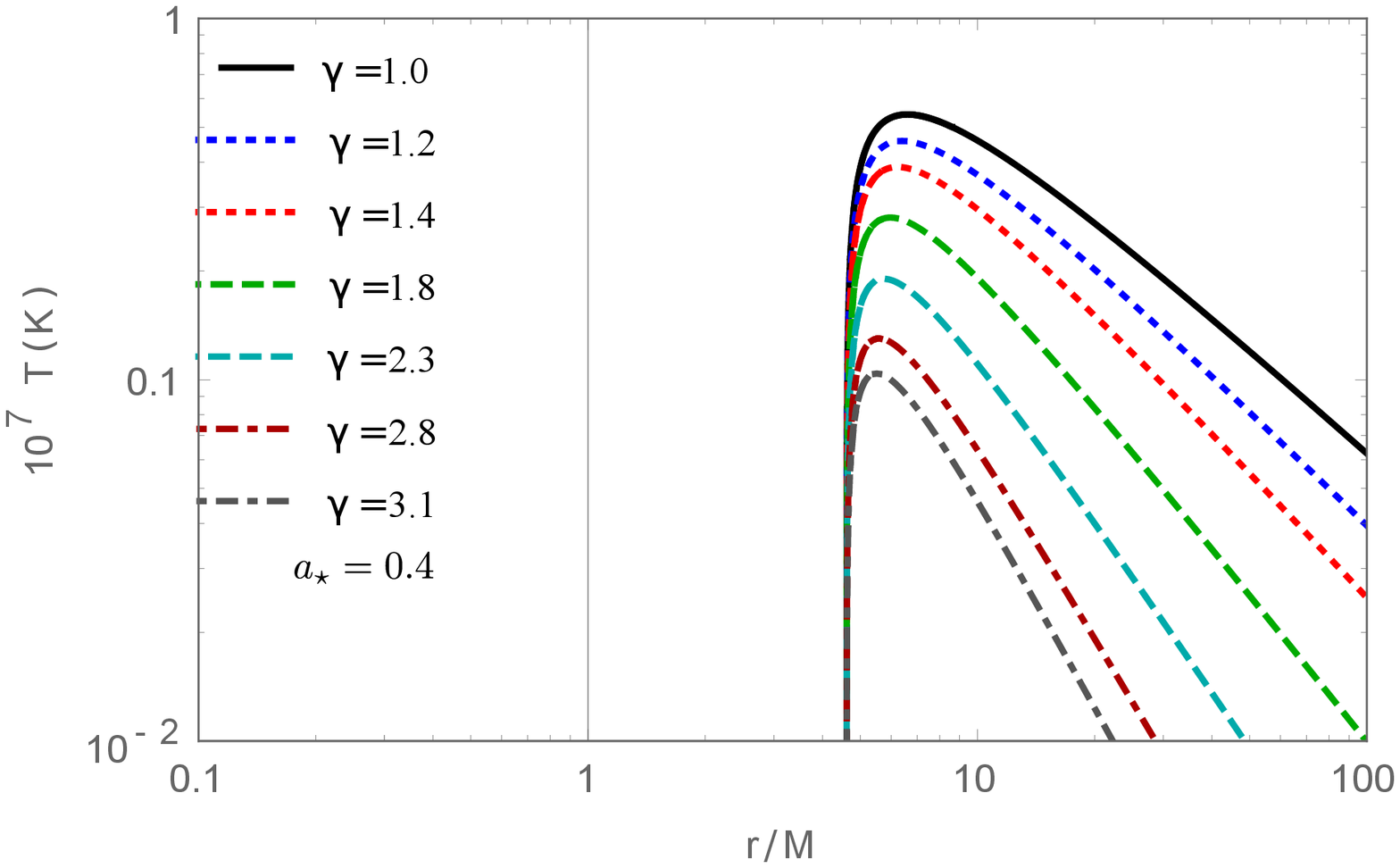}%
\newline
\noindent  \vspace{0.2cm}\newline
\includegraphics[scale=0.47]{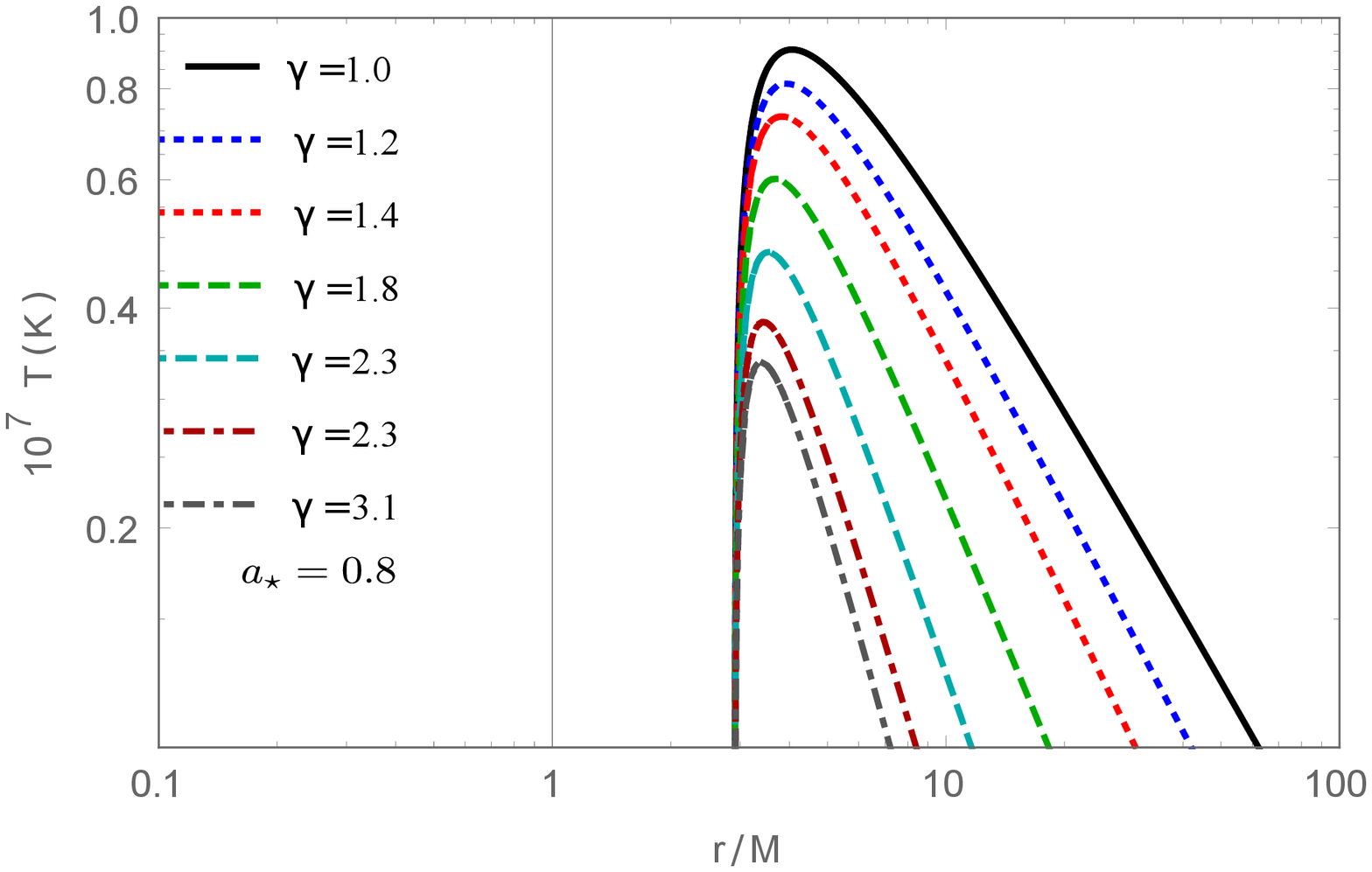}~\includegraphics[scale=0.47]{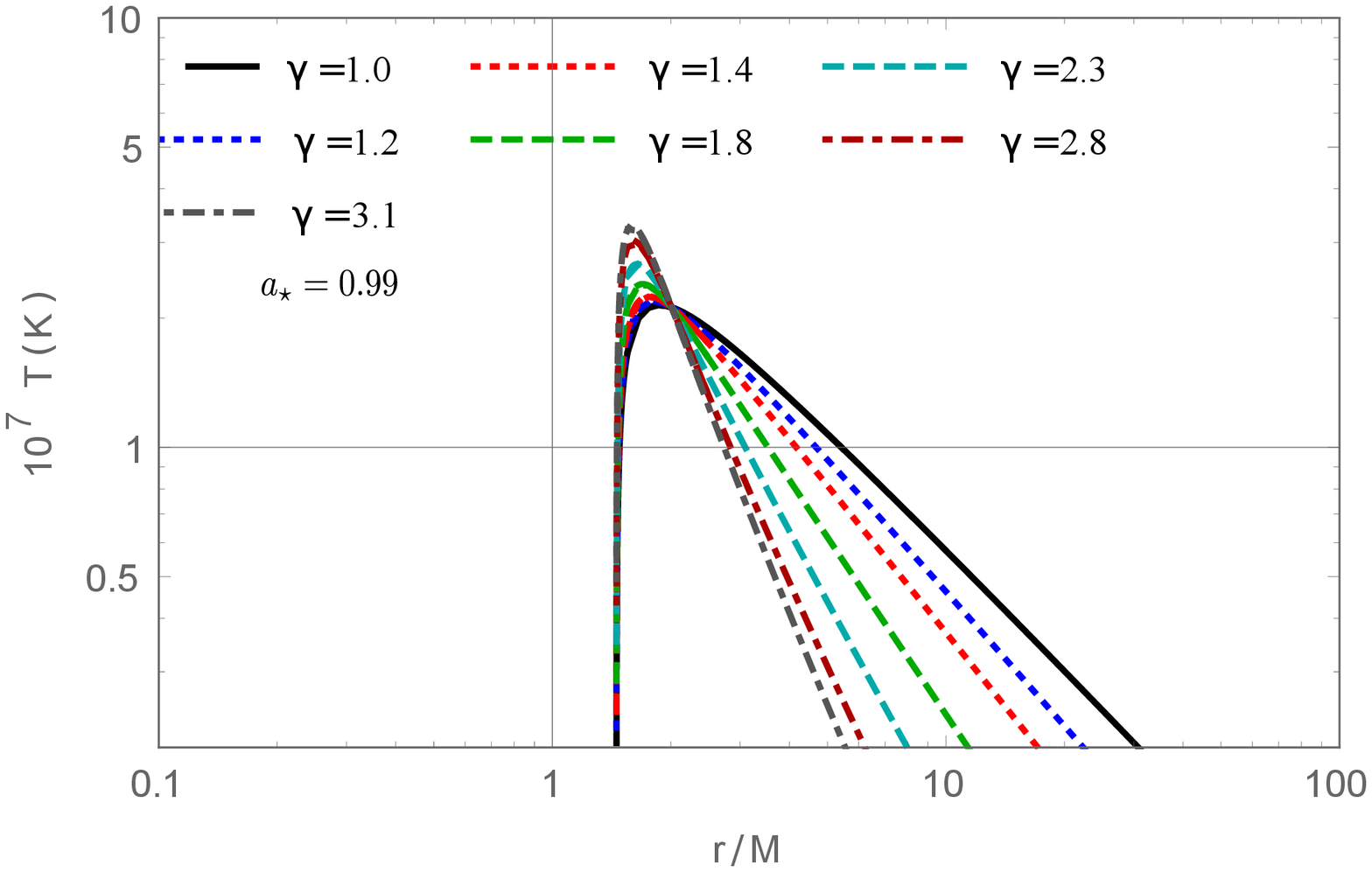}
\newline
\noindent  \vspace{0.2cm}\newline
\includegraphics[scale=0.47]{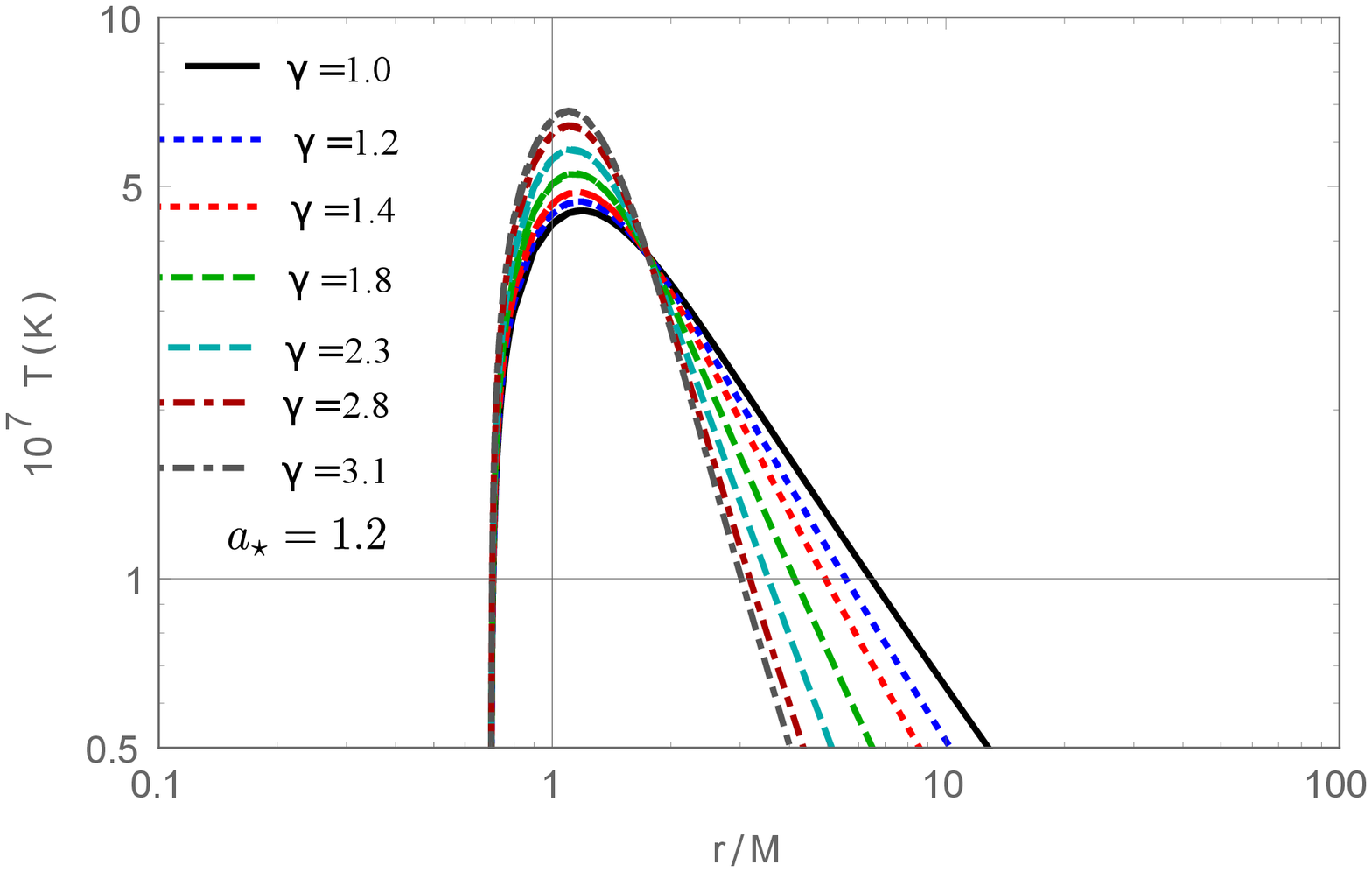}~\includegraphics[scale=0.47]{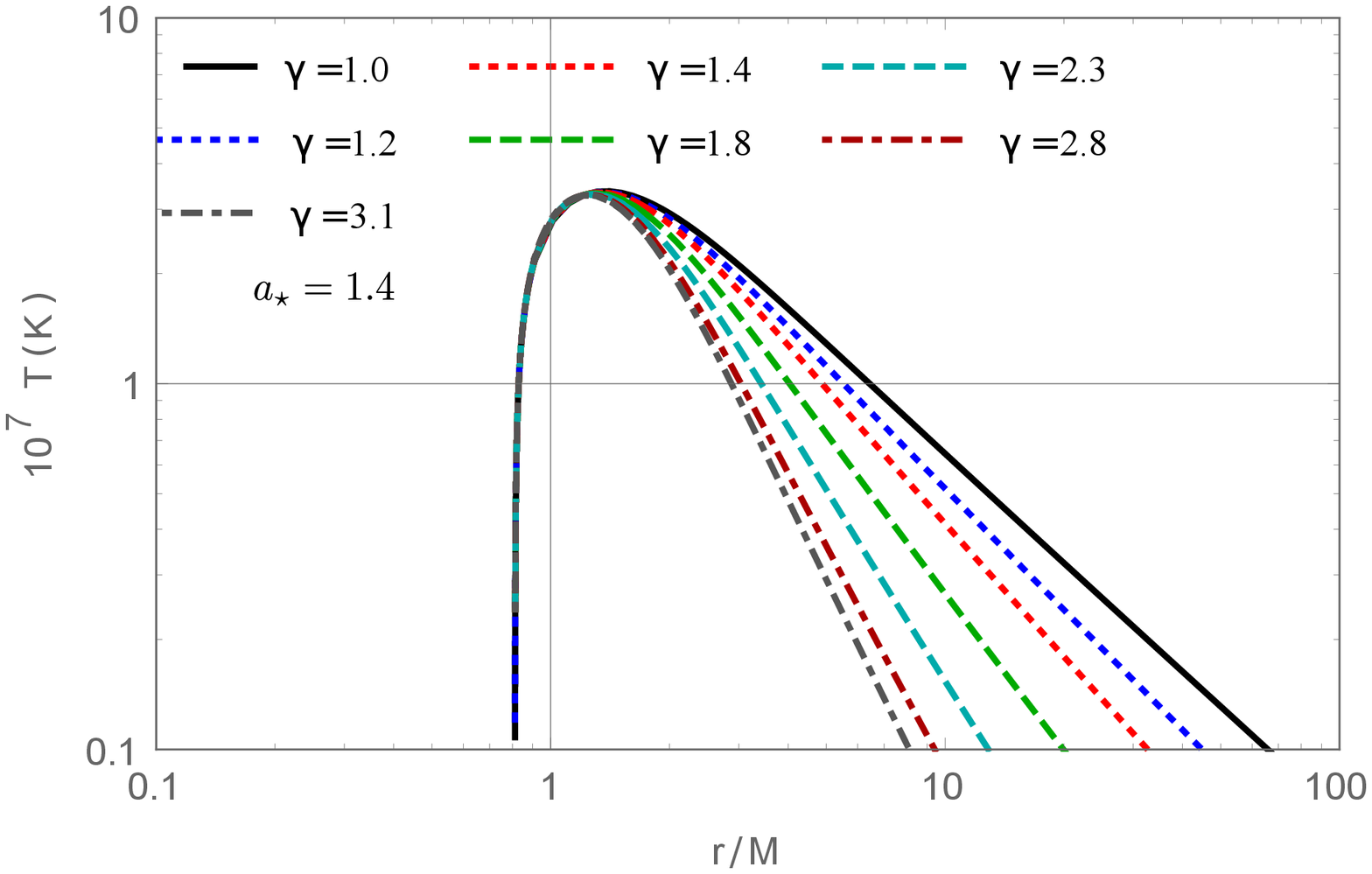}
\caption{Log-log plot of the temperature distribution $T(r)$ in the accretion disk with respect to $r/M$ for $a_\star=0,\;0.4,\;0.8,\;0.99,1.2,1.4$, and  for $\gamma=1$ (Kerr black hole for $a_\star\leq1$, and naked Kerr singularity for $a_\star>1$), $\gamma%
=1.2,\;1.4,\;1.8$ (non-trivial black holes) and $\gamma=2.3,\;2.8,\;3.1$ (naked singularities), respectively. }\label{fig4}
\end{figure*}

In Fig.~\ref{fig4} we have plotted the temperature distribution of the radiation emitted
from the disk for the same $\gamma$ and $a_\star$ values as in the previous case.  Generally, the disk temperature shows a similar dependence on the parameters $\gamma$ and $a_{\star}$ as $F(r)$ does. In the static and slowly rotating cases the disk temperature reaches its highest values in the Kerr geometry.  With increasing $\gamma$, and increasing $a_{\star}$,  the temperature profiles become much sharper, with their maxima shifting towards the inner edge of the disk.  The configurations with lower spin generate temperature profiles similar in shape to those obtained for the Kerr black holes, but with significant quantitative differences with respect to the positions and values of the maximum temperatures. In the low spin limit the disk must be cooler  as compared to the typical disk temperatures obtained for Kerr black holes, with the same spin values. Nevertheless, in the extreme spin limit, the temperature of the disk for the nontrivial black hole and naked singularity configurations will exceed the Kerr values, indicating a significant increase in the disk temperature near its inner edge, and an accentuate sharpening of the temperature profile.

The differences in the disk temperature distributions are even more important in the case of the comparison of the Kerr naked singularities with $a_*>1$ and the Brans-Dicke-Kerr type naked singularities and black holes, respectively, presented in the bottom panels of Fig.~\ref{fig4}. The maximum of the disk temperature is shifted towards the central singularity, and it shows a significant increase as compared to the case of the maximally rotating Kerr black hole. The Kerr naked singularity has the lowest maximum disk temperature, and around the inner edge of the disk the temperature of the Brans-Dicke-Kerr disks is much higher. However, the rate of the temperature decrease is different for the different types of compact objects. While for the  Brans-Dicke-Kerr type objects there is a fast decrease in the disk temperature, indicating cooler outer regions, the decrease of the disk temperature for the Kerr naked singularity takes place at a lower rate, resulting in a hotter disk at large distances from the central singularity. The maximum temperature of the disk slightly decreases with increasing $a_*$, due to the increase of $r_{ms}$.

This is a distinct observational signature that may provide the observational possibility of distinguishing between different classes of Kerr-Brans-Dicke type objects, and standard general relativistic black holes.

\subsubsection{The luminosity of the disk}

\begin{figure*}[tbp]
\centering
\includegraphics[scale=0.47]{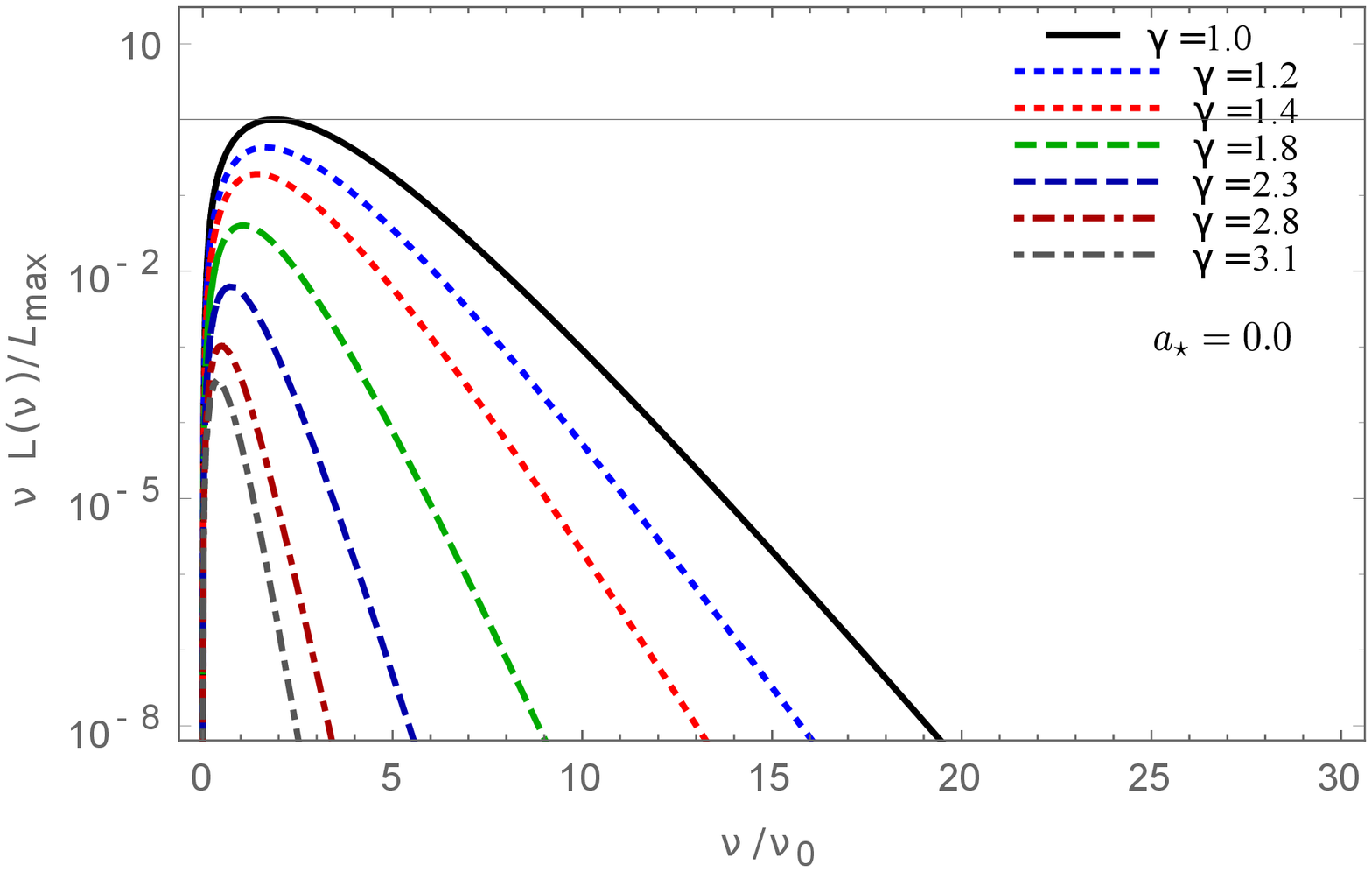}~\includegraphics[scale=0.47]{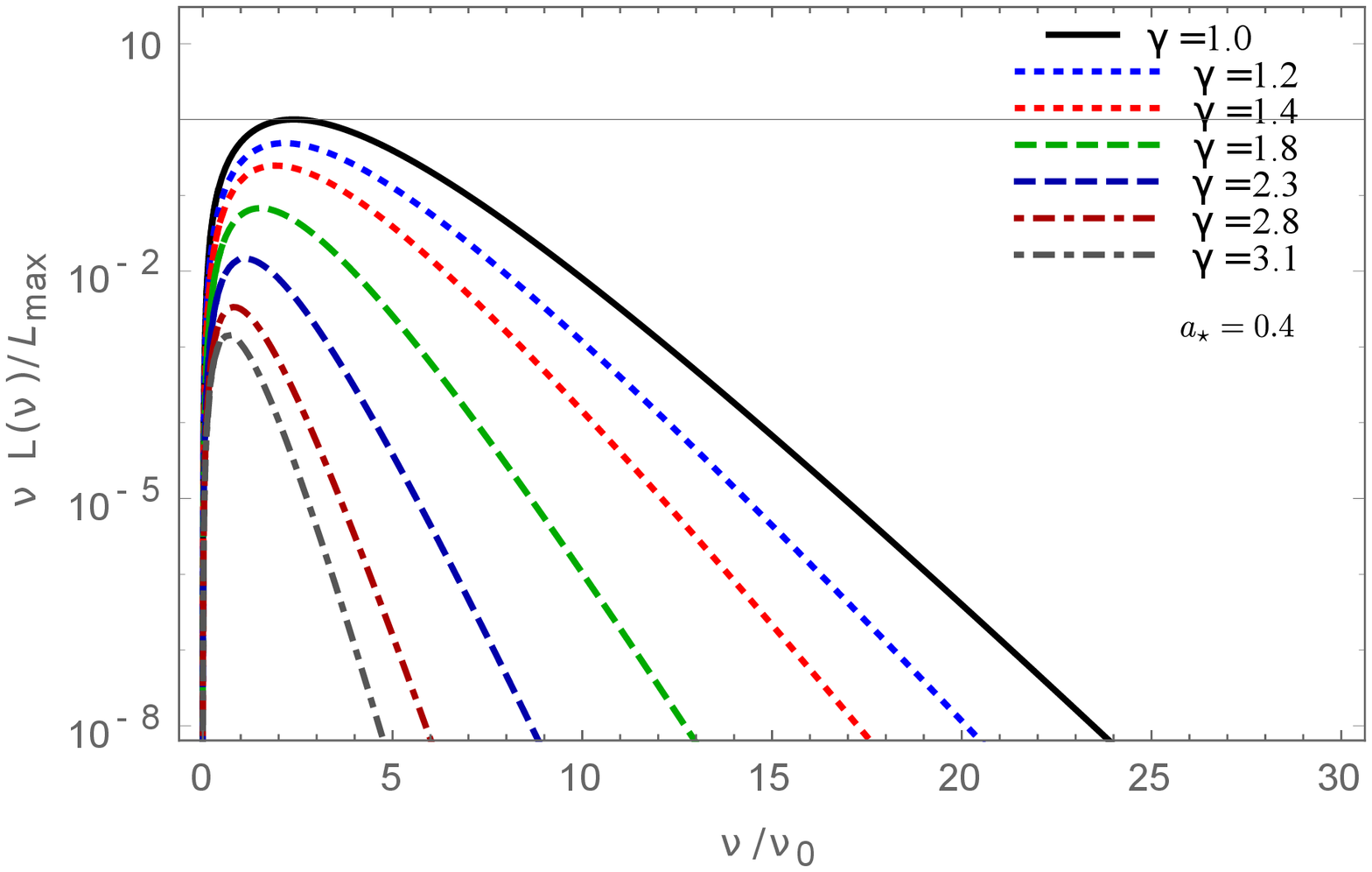}%
\newline
\noindent  \vspace{0.2cm}\newline
\includegraphics[scale=0.47]{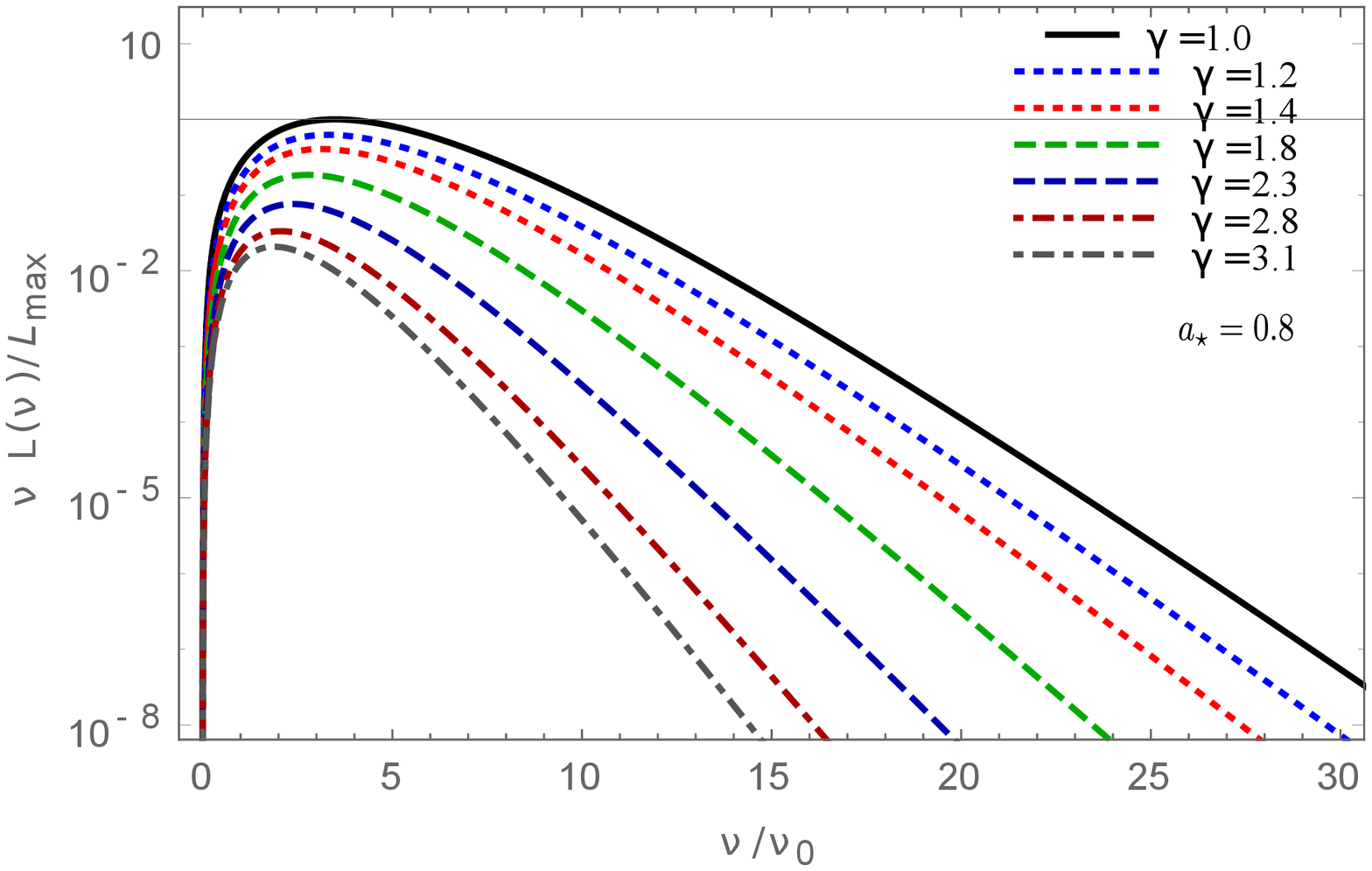}~\includegraphics[scale=0.47]{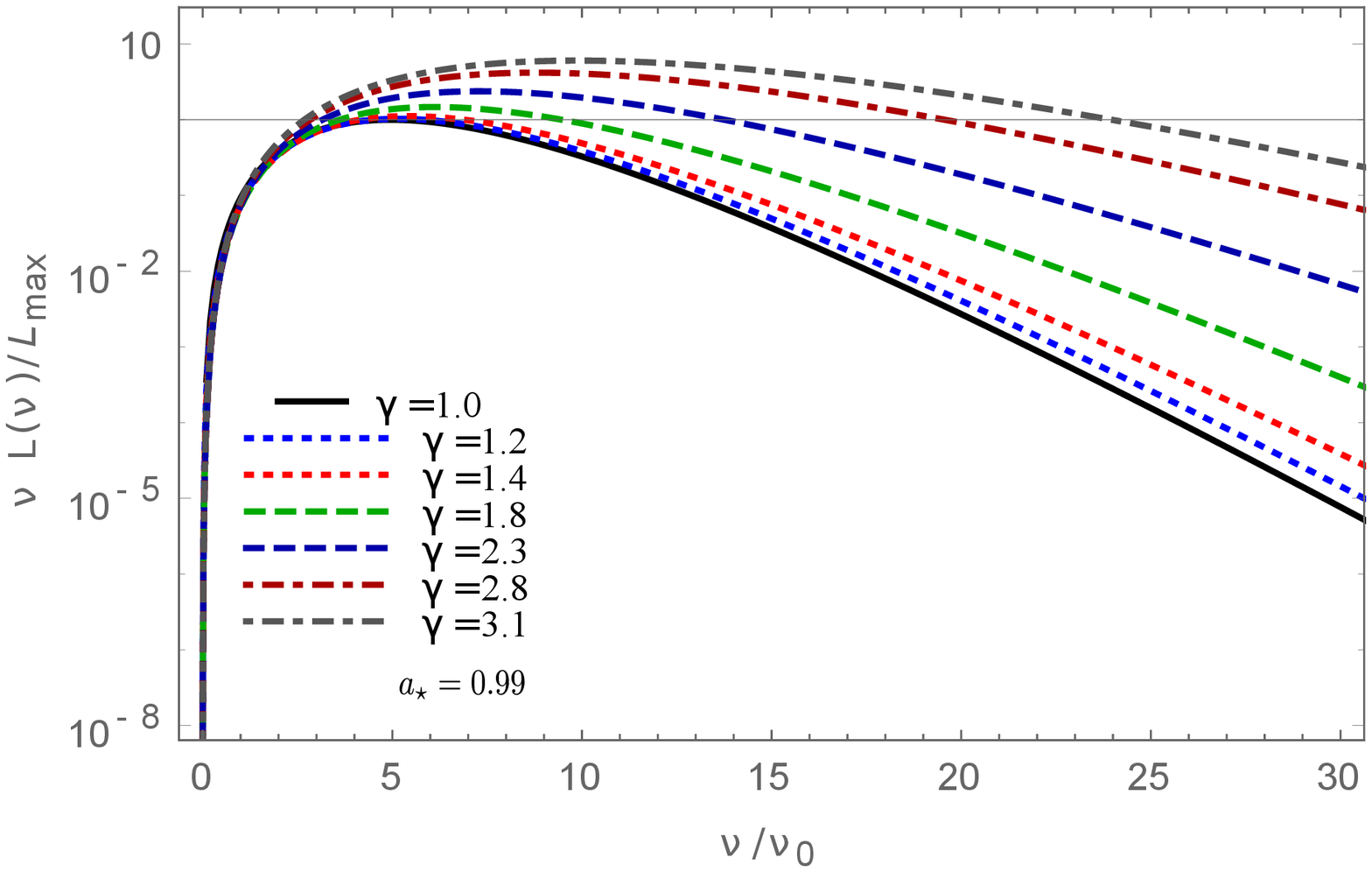}\noindent  \vspace{0.2cm}\newline
\includegraphics[scale=0.47]{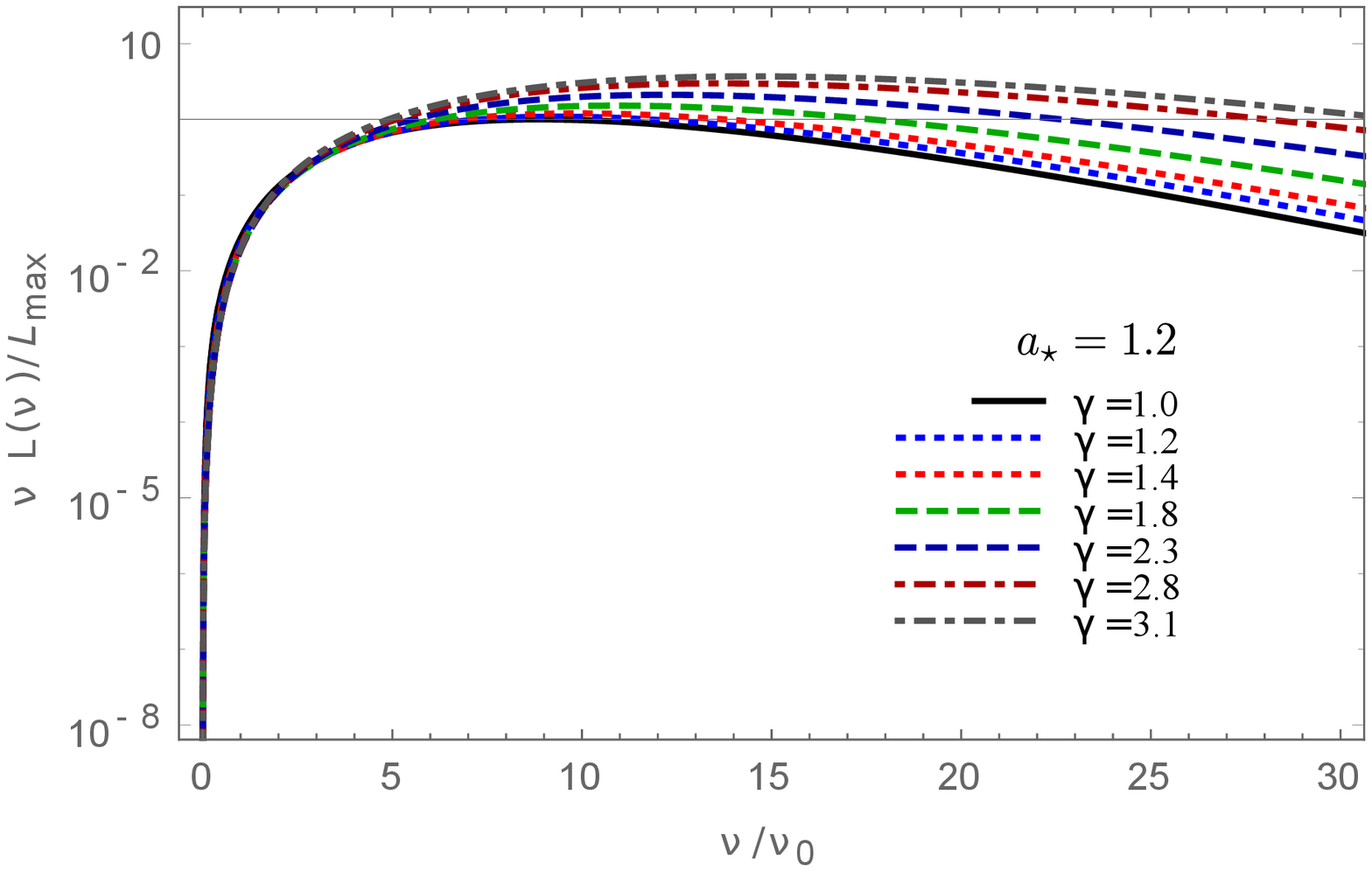}~\includegraphics[scale=0.47]{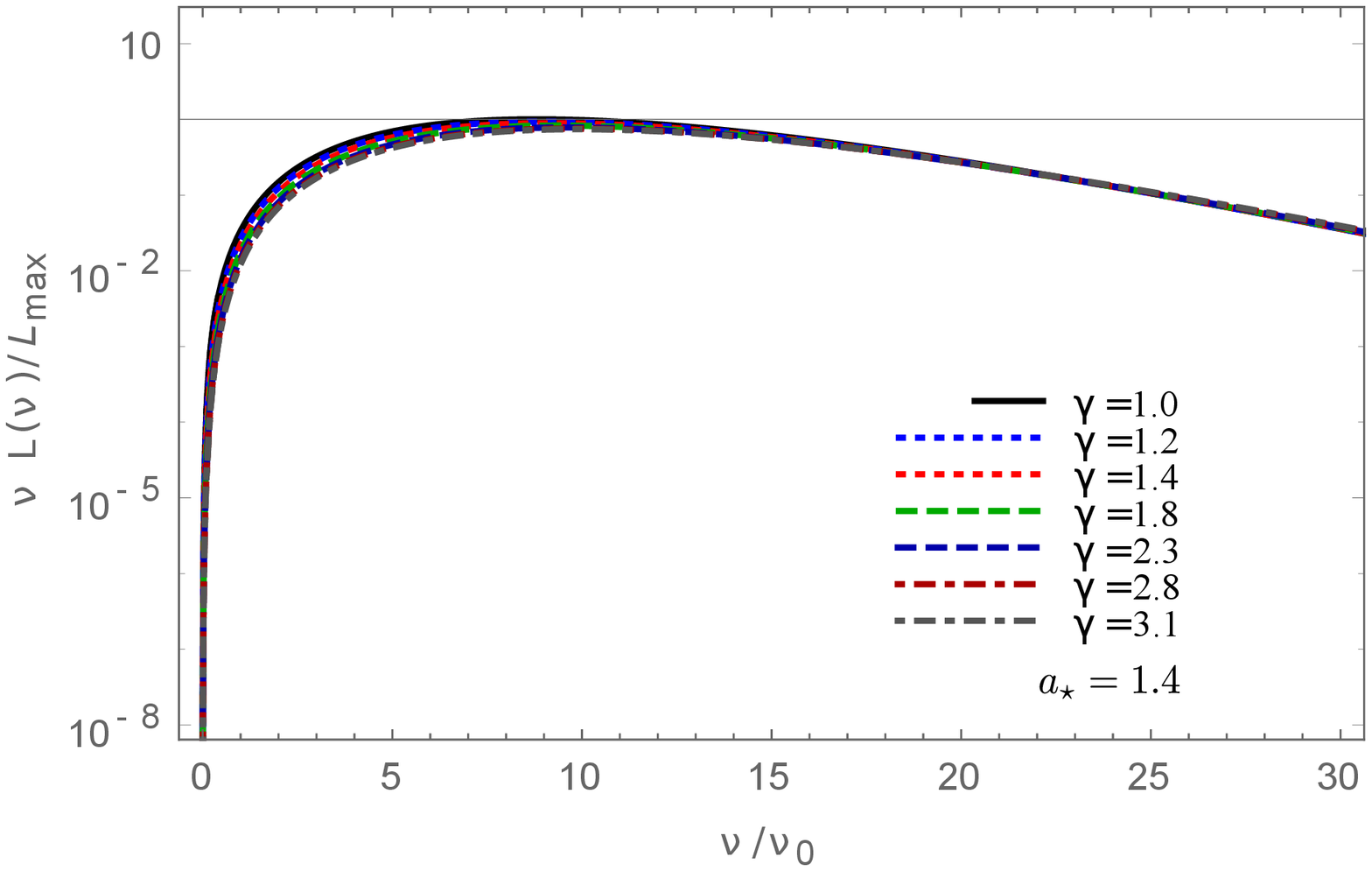}
\caption{Log plot of the normalized luminosity $\nu L(\nu%
)/L_{max}$ with respect to $\nu/\nu_0$ for $a_\star=0,\;0.4,\;0.8,\;0.99,1.2,1.4$, and  for $\gamma=1$ (Kerr black hole for $a_\star\leq1$, and naked Kerr singularity for $a_\star>1$), $\gamma%
=1.2,\;1.4,\;1.8$ (non-trivial black holes) and $\gamma=2.3,\;2.8,\;3.1$ (naked singularities), respectively. Here $%
\nu_0=2\times10^6 Hz$ and $L_{max}$ is the luminosity of the Schwarzschild disk, corresponding to $\gamma=1$, and $a_\star=0$, respectively.}\label{fig5}
\end{figure*}

In Fig.~\ref{fig5} we have plotted, for the same set of values of the parameters $a_*$ and $\gamma$, the normalized luminosity $\nu L(\nu)/L_{max}$, as a
function of the frequency for different values of $\gamma$ and $a_\star$, which were calculated from the luminosity equation Eq.~(\ref{L}). For $L_{max}$ we have adopted the luminosity of the Schwarzschild disk, with $\gamma=1$, and $a_\star=0$, respectively. As expected, the same features observed in the behavior of the energy flux distribution $F(r)$, and of the disk temperature in the black hole and naked singularity geometries, are present in the luminosity distributions. For slow rotations, the Kerr luminosity of the disk exceeds by almost three orders of magnitude the luminosity of the Kerr-Brans-Dicke naked singularities. Moreover, the maximum of the spectra is shifted towards higher values of $\nu/\nu_0$ in the disk, and this effect is significant in the case of naked singularities. Generally, the maximal amplitudes increase with the increase of the spin parameter $a_{\star}$, that is, the accretion disks of both black holes and naked singularities become hotter by rotating faster. The fast rotation leads to a blueshifted surface radiation, with higher intensity. Still, even in the slow rotation case the disk spectra exhibit important differences between black holes and naked singularities.

For fast rotation ($a_{\star}=0.99$), the distribution of the luminosity completely changes, with both the luminosity of the non-trivial black holes and of the naked singularities exceeding the Kerr luminosity. In this case there is a shift with respect to the Kerr maximum towards the outer edge of the disk, with the accretion disk becoming much hotter in the areas distant with respect to the singularity. Hence the maximal amplitudes of the spectra of the non-trivial black holes and naked singularities have much higher values then in the case of the Kerr black hole disk spectra, indicating that the spectral properties of the disks are very sensitive to the variations in the spin at high rotation speeds.  Thus the relative shifts in the cut-off frequencies and the spectral maxima for extreme black hole and naked singularity geometries may provide another tests for discriminating the Kerr black holes and non-trivial black holes, and naked singularities, respectively.

The behavior of the disk luminosity significantly changes in the case of the Kerr naked singularities and of the Brans-Dicke-Kerr naked singularities and black holes respectively, corresponding to $a_*>1$. These cases are presented in the bottom panel of Fig.~\ref{fig5}. For $a_*=1.2$, the luminosity of the Kerr naked singularity is significantly lower as compared to the luminosity of the Brans-Dicke-Kerr naked singularities/black holes. In all cases the maximum luminosity of the disk is reached in its inner regions, with the location of the maximum in the frequency spectrum approximately equal to the case of the maximally rotating Kerr black hole (around $5\nu/nu_0$). However, the distribution of the  luminosity on the disk is different as compared to the maximally rotating Kerr black hole case. The rate of decrease of the luminosity with increasing radiation frequency is fastest for the Kerr naked singularity, indicating a higher luminosity of the outer regions of the disk for higher radiation frequencies for Brans-Dicke-Kerr naked singularities/black holes. An interesting situation appears in the luminosity behavior for increasing $a_*$. For $a_*=1.4$ (right plot in the bottom panel of Fig.~\ref{fig5}) the differences in the luminosities of the Kerr naked singularities and of the Brans-Dicke-Kerr naked singularities/black hole become negligible, and basically a unique spectrum describe the frequency dependence of the luminosity of the disk. However, this spectrum differs from the one corresponding to the maximally rotating Kerr black hole, and the corresponding Brans-Dicke-Kerr objects, with a much slower rate of decrease of the disk luminosity with increasing frequencies.

\subsection{Eddington luminosity of the disk}

For the case of a boson star, an interesting effect, involving the Eddington luminosity, was discussed in \cite{To02}. The Eddington luminosity, representing from a physical point of view the limiting luminosity that can be obtained from the equality of the attractive gravitational force and of the repulsive radiation force, is given by
\be
L_{Edd}=\frac{4\pi Mm_p}{\sigma _T}=1.3\times 10^{38}\left(\frac{M}{M_{\odot}}\right)\; {\rm erg/s}.
\ee
 On the other hand since the mass distribution of the bosonic field forming a boson star has a radial distance dependent mass distribution, with $M=M(r)$, it follows that for bosonic systems the Eddington luminosity becomes a spatial coordinate dependent quantity, so that $L_{Edd}(r)\propto M(r)$.

A similar effect occurs for the case of the Kerr-Brans-Dicke solutions considered in the present study.  One can associate to the Brans-Dicke scalar field, described by its energy-momentum tensor, a mass distribution $M_{Edd}(r)$ along the equatorial plane of the disk, given by
\be\label{mass}
M_{Edd}(r)^{(\varphi)}=-4\pi\int_{r_s}^r T^{\varphi0}_{~~0}r^2 dr
=2\pi\int g^{rr}\varphi_{,r}\varphi_{,r}r^2 dr.
\ee

Then the Eddington luminosity of the scalar field can be obtained as
\be
L_{Edd}^{(\varphi)}(r)=\frac{4\pi M_{Edd}(r) m_p}{\sigma_T}=1.3\times10^{38}\f{M_{Edd}(r)}{M_\odot} \f{\textmd{erg}}{\textmd{s}}.
\ee

 Using Eq.~\eqref{mass}, we obtain the Eddington luminosity associated to the Kerr-Brans-Dicke scalar field as
\begin{align}
L_{Edd}^{(\varphi)}(\rho)&=\frac{32\pi^2 m_p}{\sigma_T}M\int_{\rho_{ms}}^\rho \f{|1-\gamma|(1-\rho)^2d\rho}{(a_\star^2+\rho^2-2\rho)^{1+|1-\gamma|}}\nonumber\\
&=1.3\times10^{40}\frac{M}{M_\odot}l_{Edd}^{(\varphi)},
\end{align}
where $\rho=r/M$, and
\be
l_{Edd}^{(\varphi)}(\rho)=\int_{\rho_{ms}}^\rho \f{|1-\gamma|(1-\rho)^2d\rho}{(a_\star^2+\rho^2-2\rho)^{1+|1-\gamma|}}.
\ee

In Fig.~\ref{figedd} we have plotted $l_{Edd}^{(\varphi)}(\rho)$ as a function of $\rho=r/M$.

\begin{figure*}[tbp]
	\centering
	\includegraphics[scale=0.47]{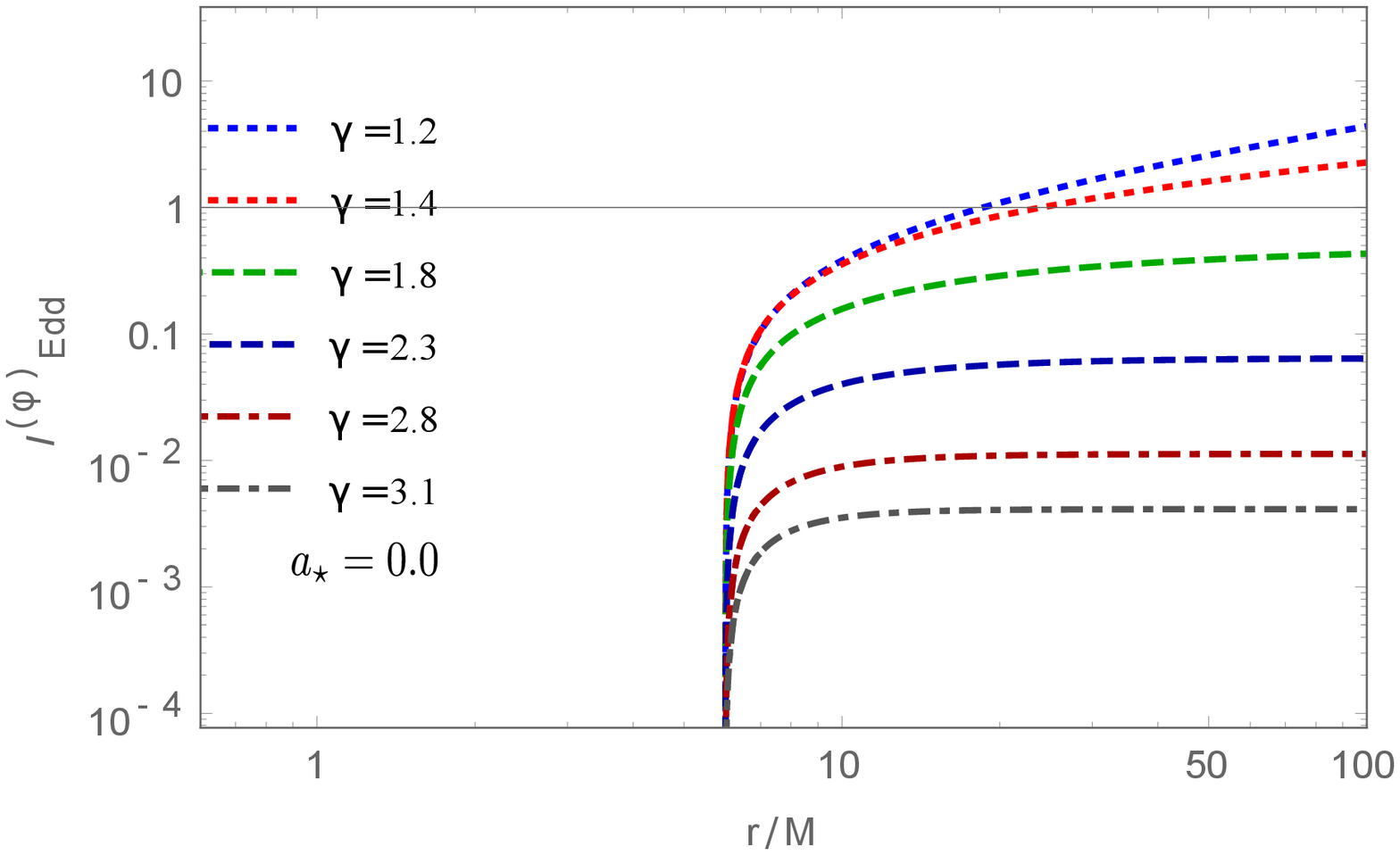}~\includegraphics[scale=0.47]{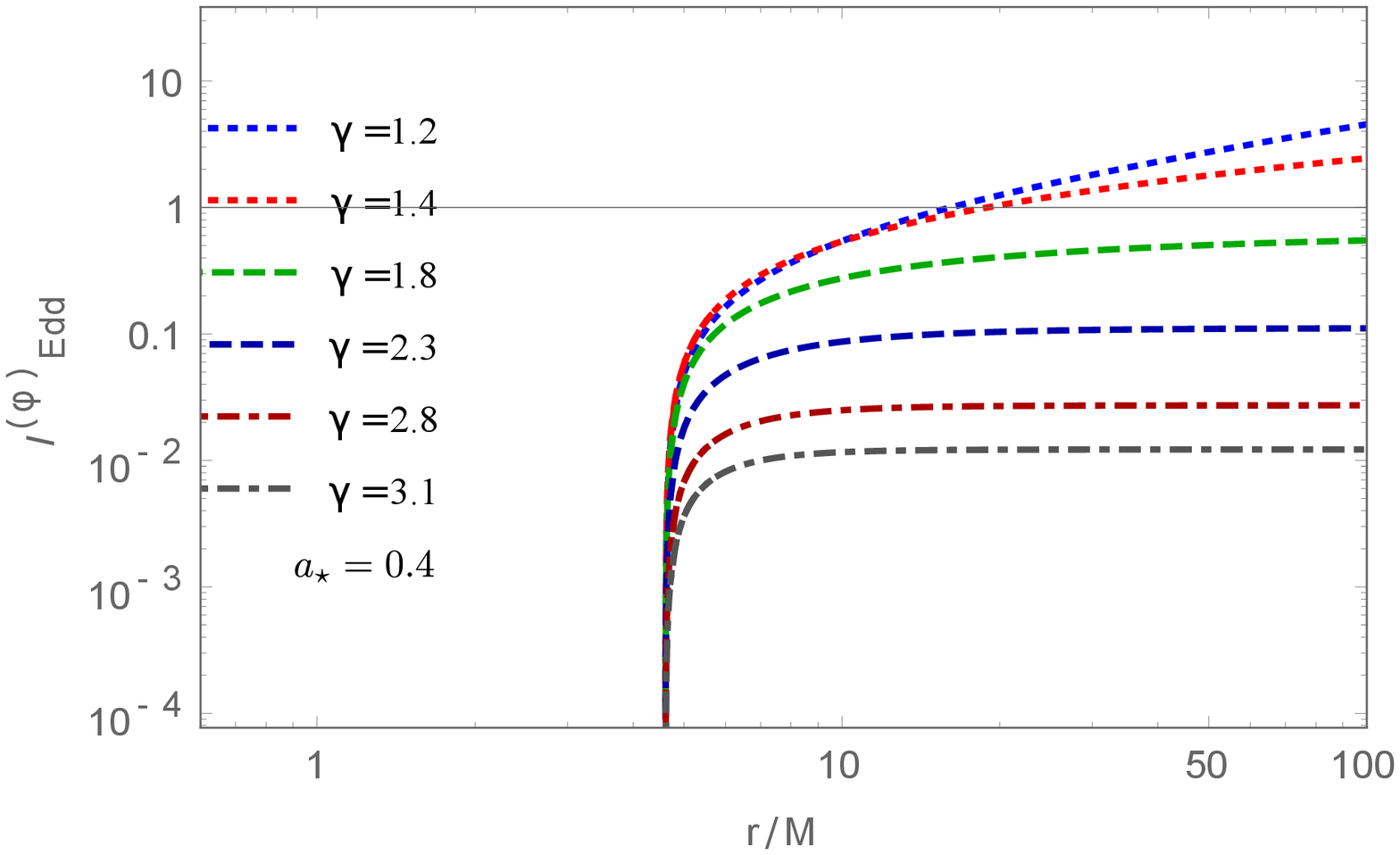}%
	\newline
	\noindent  \vspace{0.2cm}\newline
	\includegraphics[scale=0.47]{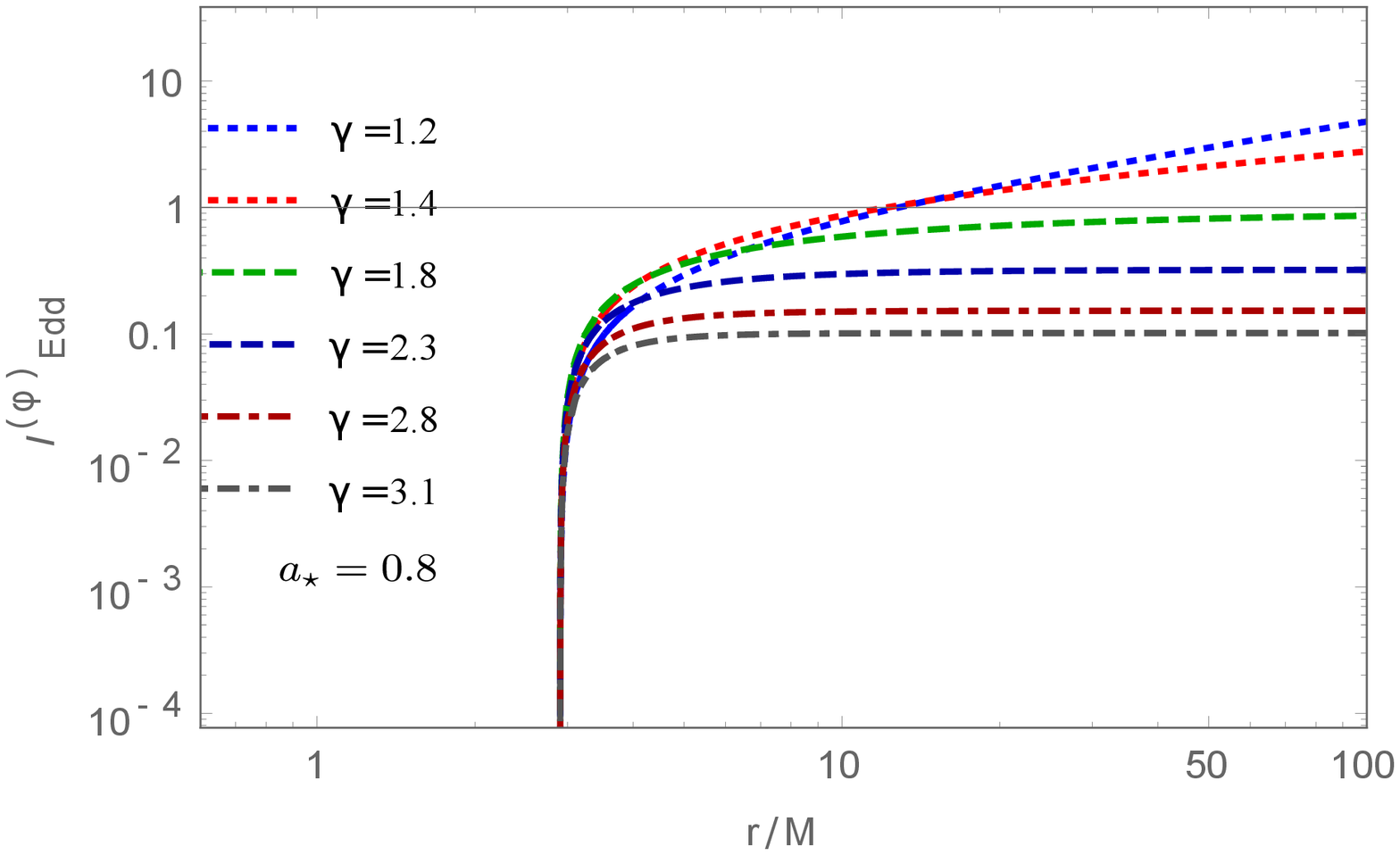}~\includegraphics[scale=0.47]{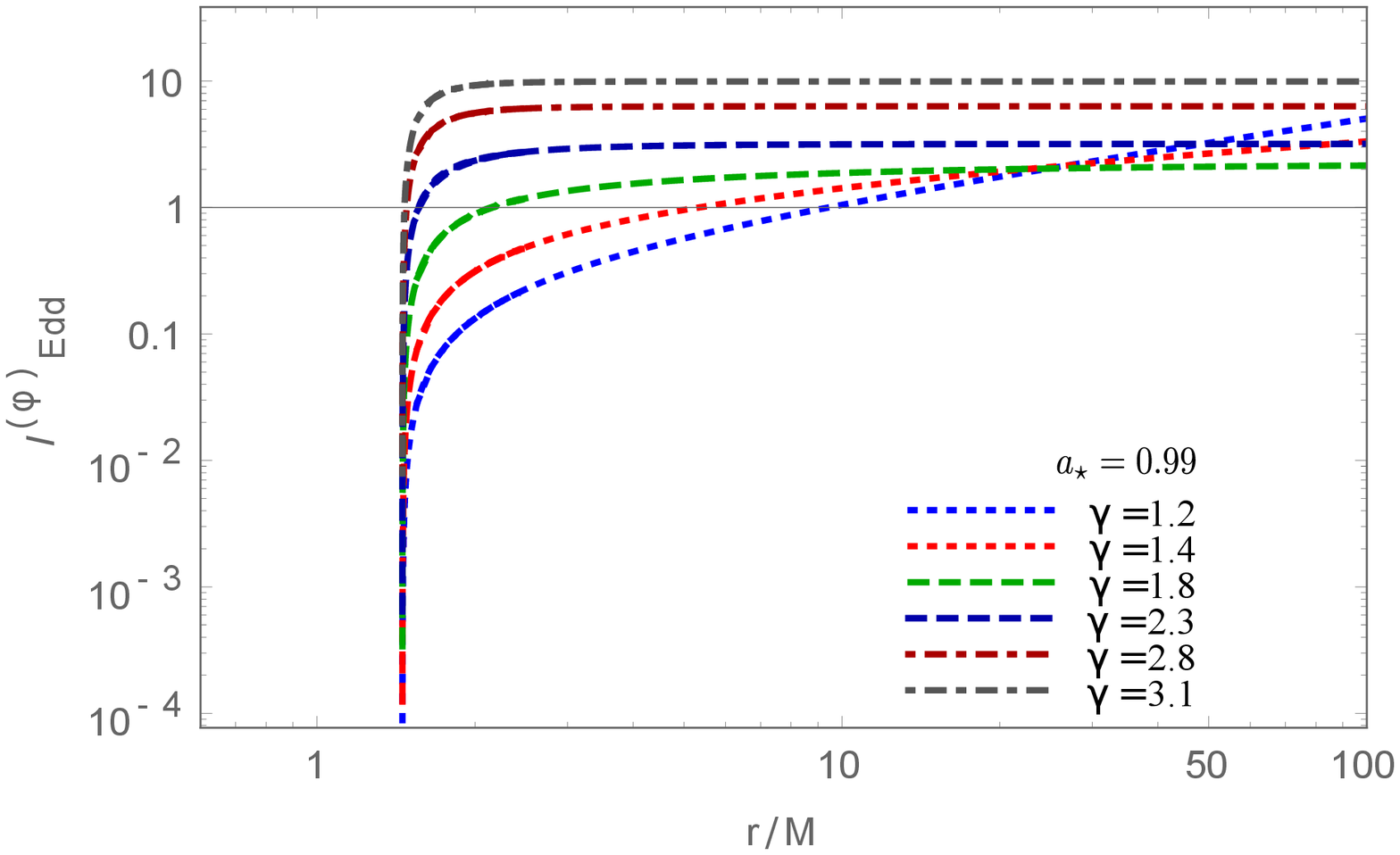}\noindent  \vspace{0.2cm}\newline
	\includegraphics[scale=0.47]{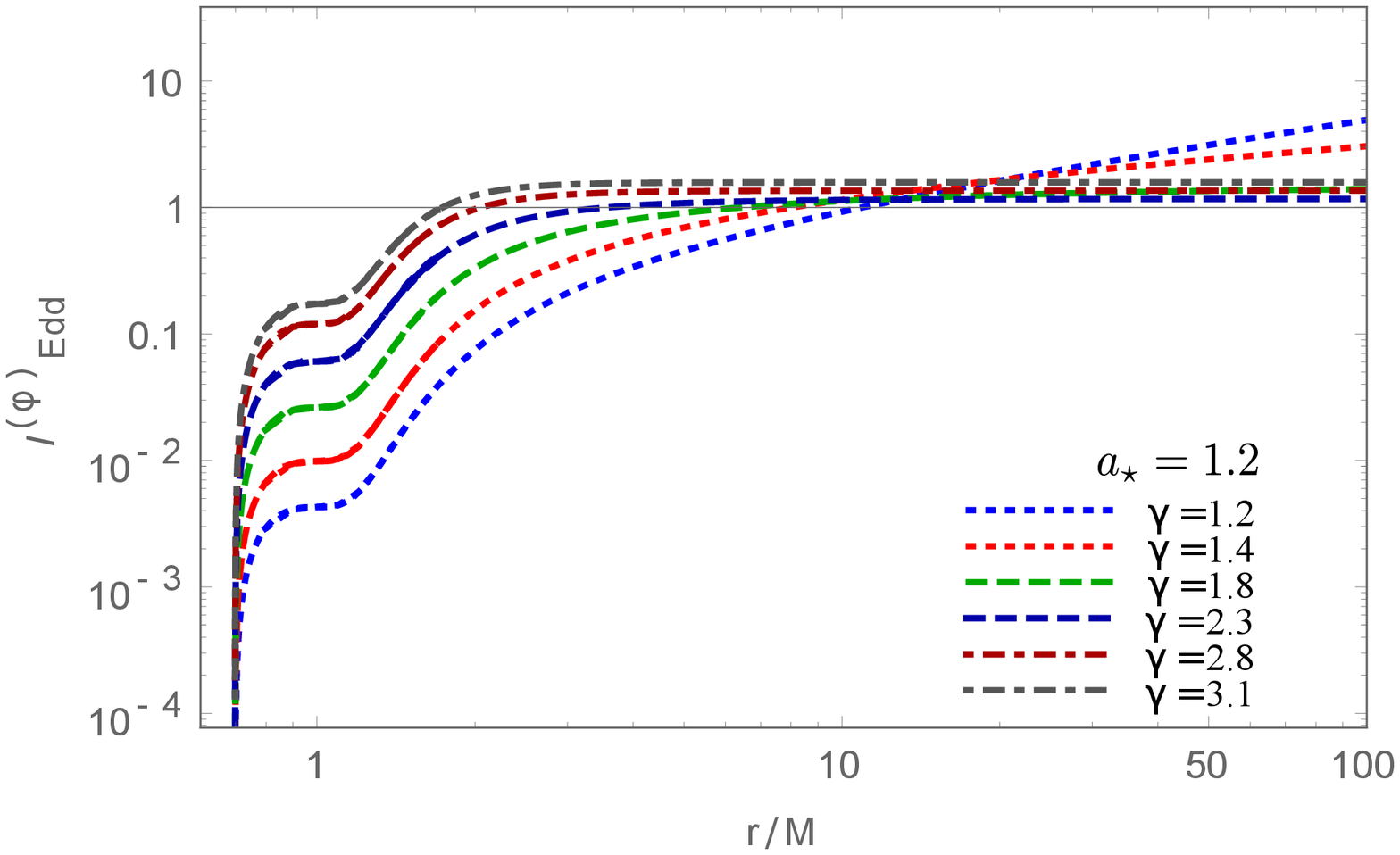}~	\includegraphics[scale=0.47]{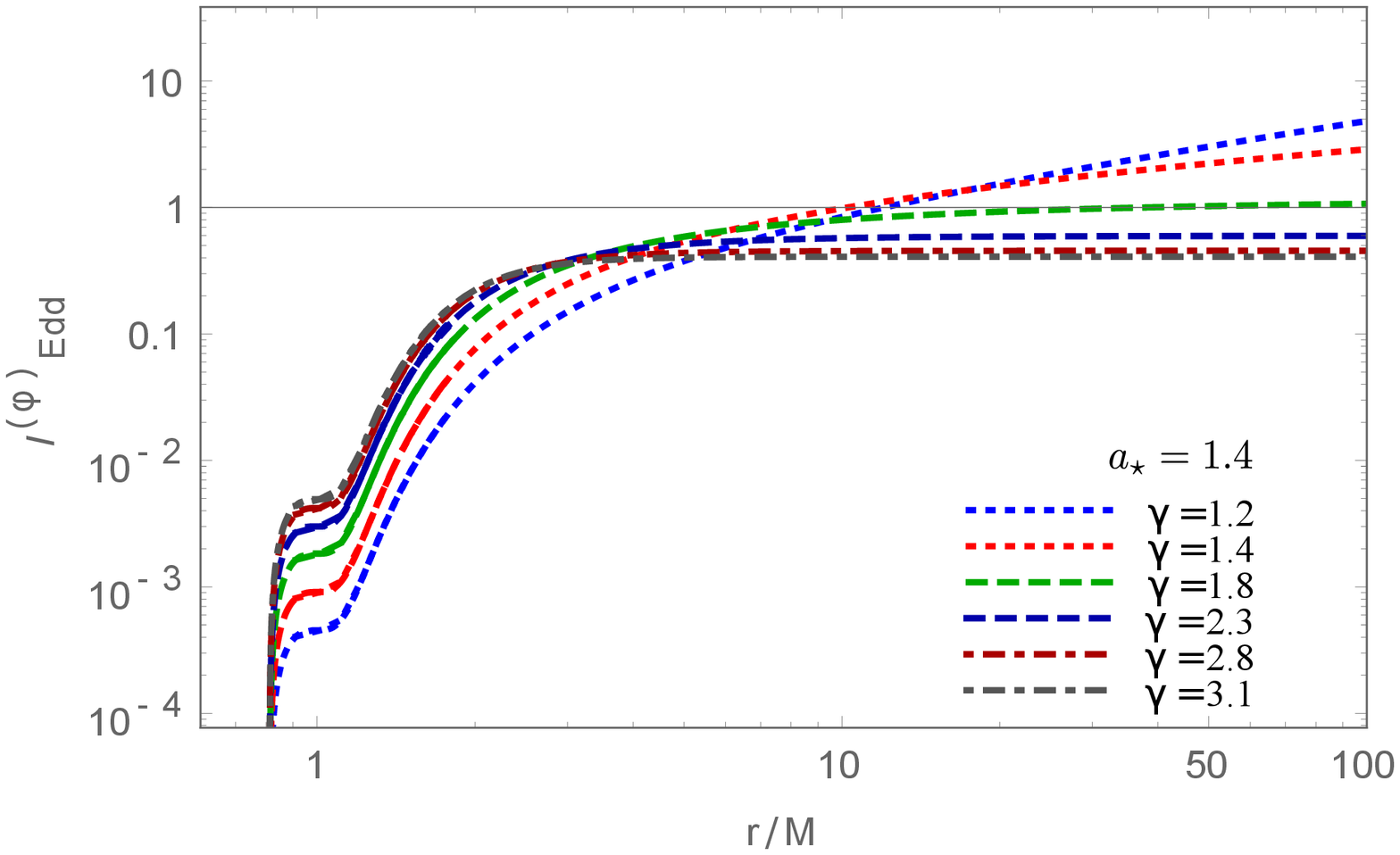}
	\caption{Log-Log plot of the normalized Eddington luminosity $l_{Edd}^{(\varphi)}(\rho)$ with respect to $r/M$ for $%
		a_\star=0,0.4,0.8,0.99,1.2,1.4$, and  for $\gamma%
		=1.2,1.4,1.8$ (non-trivial black holes) and $\gamma=2.3,2.8,3.1$ (naked singularities), respectively.}\label{figedd}
\end{figure*}

There is a significant difference in the Eddington luminosity associated to the scalar fields giving birth to black holes and naked singularities, respectively. The luminosity $l_{Edd}^{(\varphi)}(\rho)$ is higher for black holes as compared to the naked singularities, and it has a strong dependence on $a_{\star}$. There is an initial very rapid increase in the Eddington luminosity generated near the marginally stable orbit $r_{ms}/M$, followed by a plateau phase, in which $l_{Edd}^{(\varphi)}(\rho)$ is almost constant. The maximum plateau values of $l_{Edd}^{(\varphi)}(\rho)$ depend strongly of the type of the central object (black hole or naked singularity), and they are higher for black holes. The Eddington luminosity of Brans-Dicke-Kerr type scalar field increases with increasing spin $a_{\star}$, and it extends to very high distances from the central object.

The differences in the behavior of the Eddington luminosity between different Brans-Dicke-Kerr type objects become even more significant for $a_*>1$, as one can see from the bottom panels of Fig.~\ref{figedd}. The Eddington luminosity still increases with increasing $\gamma$, but their numerical values are generally smaller than for the $a_*=0.99$ case. $l_{Edd}^{(\varphi)}(\rho)$ increases very rapidly near $r_{ms}$ with increasing $r/M$, and shows a complex behavior at the inner edge of the disk. For larger values of $\gamma $, corresponding to naked singularities, it reaches very quickly a plateau phase, becoming a constant at around $r/M\approx 5$. For smaller values of $\gamma $ (describing the non-trivial black hole solutions) the plateau phase appears at much higher radii, with $l_{Edd}^{(\varphi)}(\rho)$ slowly increasing along the disk. The maximum values of $l_{Edd}^{(\varphi)}(\rho)$ tend to decrease with increasing $a_*$ and $\gamma$. Hence there is a significant decrease in the Eddington luminosity of Brans-Dicke-Kerr naked singularities as compared to the one of the non-trivial black holes.

\section{Discussions and final remarks}\label{sect5}

In the present paper we have presented a comparative analysis of the properties of the accretion disks that could form around black holes, massive objects possessing an event horizon, and naked singularities, hypothetical theoretical general relativistic objects, characterized by the absence of an event horizon, and a central singularity, respectively. For our study we have considered a rotating solution of the Einstein - massless scalar field equations \cite{sol}, which has the advantage of containing in a single metric form three distinct types of objects, corresponding to different choices of the model parameter $\gamma$. For $\gamma =1$  the solution reduces to the standard Kerr black hole of general relativity. Naked singularity type solutions are obtained in the range $\gamma \in \left(2,\infty\right)$, while for $\gamma \in (0,1)$ and $\gamma \in (1,2)$ we obtain non-trivial black hole solutions, characterized by the presence of an event horizon, and with physical and geometrical properties different from the Kerr black hole properties.
As a first step in any study on accretion disk properties one must investigate the motion of the massive test particles in the gravitational potential of the  central massive object. The characteristics of the motion depend on the  values of the mass, spin parameter, and model parameter, respectively. The positions of the marginally stable orbits, photon orbits and of the marginally bound orbits are determined by the $g_{tt}$, $g_{\phi \phi}$ and $g_{t\phi}$ components of the metric tensor, which in the case of the Brans-Dicke-Kerr solution coincide with their standard general relativistic counterparts. Hence the geometric characteristics of disks in the scalar field Brans-Dicke-Kerr geometry are the same as in general relativity, and all the marginally stable orbits
that are located outside the naked singularities and the black holes. Therefore the particles in the disk cannot reach, and be in direct contact with the singularity in an equilibrium configuration. The frame dragging properties of Brans-Dicke-Kerr  naked singularity are also identical with the Kerr and non-trivial black hole cases. Moreover, the conversion efficiency of the accreting mass into radiation of naked singularities and black holes in the Brans-Dicke -Kerr is  identical to the standard Kerr case, and none of the considered objects could provide a larger mass - radiation conversion efficiency
than the Kerr black holes.

However, the above points do not imply that the disk properties for the different types of black holes and the naked singularities of the Brans-Dicke-Kerr theory are identical. Due to the differences in the expression of the determinant of the metric tensor the main physical properties of the disk are dependent of the exterior geometry of the central object. In all covariant general relativistic formulations of disk models the thermodynamic quantities are obtained by integrating over the invariant four volume element. The behavior of the volume element depends on the type of the central object (black hole or naked singularity), and near the inner edge of the disk it gives the dominant contribution to the emitted flux, and the temperature and spectrum.  Therefore, the properties of the disk radiation are significantly different for black holes and naked singularities. Generally, even that on a qualitative level there are many similarities between the Kerr or non-trivial black hole disks, and the naked singularity disks, the thermodynamic/electromagnetic properties of the naked singularities could differ significantly quantitatively (by several orders of magnitude) from the non-trivial or Kerr black hole disks.

In our investigations of the black hole and naked singularity properties we have used the thin disk model, which is an obviously idealized physical model, built upon several simplifying physical and geometrical assumptions \cite{PaTh74, KoHa10}. In particular,  the self-gravity of the disk is neglected, and it is assumed that the disk is located in the central plane of the massive object. The most important assumption is that the disk is geometrically thin, and one can neglect its vertical size. Any change in the parameters of the central object during a small time interval $\Delta t$ is neglected, but this time interval is considered to be large enough for measuring the total inward energy and mass flows at any point in the disk. From a mathematical point of view we have assumed
that the energy-momentum  tensor of the disk matter can be algebraically decomposed with respect to its four-velocity. Moreover, in our approach the averaged dynamics of the baryons over the azimuthal angle and $\Delta t$ is given by the circular geodesic motion in the equatorial plane. From a physical point of view we have assumed that the heat flow within the disk in the radial direction is negligible, and it is important only in the vertical direction.
The energy is carried to the disk surface by thermal photons, and the photons are emitted on average only in the vertical direction. Finally, we have neglected
the energy of the photons emitted vertically from the disk surface when studying the momentum and energy transport between the different regions of the disk. Once any of the above conditions are not satisfied, the thin disk model cannot be applied anymore. Nevertheless, since in the present model at the inner edge of the disk the variation of the volume element gives the dominant contribution to the flux, temperature, and spectrum of the disk, it turns out that this  contribution is much larger than the effects on the thermodynamical parameters of the disk that could result from some theoretically improved disk models.

The possibility of distinguishing black holes and naked singularities via their accretion disk properties was investigated in detail in \cite{ns23,ns24} and \cite{ns25}, respectively. A basic difference between the present approach and the investigations in \cite{ns24,ns25}  is that in these papers the authors consider magnetized accretion disks in the Kerr geometry only \cite{ns24, ns25}, while in our study we consider a different geometry, and the effects of the magnetic field are ignored. The presence of the magnetic fields strongly affects the orbital motion of the particles, which influences the emitted flux through the modification of the innermost stable orbits. An interesting particularity of the Kerr-Brans-Dicke metric, used in the present study, is that the particle motion in the disk is the same as in the standard Kerr metric. In \cite{ns25} the authors consider emission from hot spots on the disk, and they show that the emission from a hot spot orbiting near the innermost stable circular orbit of a naked singularity  in a dipolar magnetic field is significantly harder than the emission of the same hot spot in the absence of such a magnetic field.  To obtain the geodesics of photons between a plane placed at the position of the observer and the surface of the disk a ray-tracing technique is developed for this specific problem.

It is generally believed that the astrophysical objects grow via accretion, and that around most of black holes and active galactic nuclei (AGN's)
there exist gas clouds surrounding the central object, forming an associated accretion disk. The gas can exist in
either the atomic or the molecular state.
The disks have very different length scales, ranging from AU-to-parsec scales in AGN's to solar radius-to-AU scale disks in protostellar objects \cite{Spruit}.  The gas clouds  form an optically and geometrically  thick torus (or warped disk).  The disk absorbs most of the soft X-rays and  the ultraviolet radiation.

The temperature distribution in accretion disks depend on the the mass accretion rate, the mass of the central black hole, and on the location of the emission point in the accretion disk, respectively. For stellar-mass black holes accreting at about 10\%
of their Eddington limit the thermal spectrum of the inner part of the accretion disk is in the soft X-ray band (0.1-1 keV), while for the supermassive black holes it is in the optical/UV band (1-10 eV) \cite{Bambi1}. Through the inverse Compton scattering by the hot electrons in the
corona, the thermal photons gain energy, and they convert into X-rays, having a characteristic power-law component. The X-ray photons illuminate the disk, generating a new, reflection component, with strong fluorescent emission lines. Usually the most noticeable characteristic of the reflection spectrum is the iron K$\alpha$ line, located, in the case of neutral or weakly ionized iron, at an energy of 6.4 keV,  and which for H-like ions shifts up to 6.97 keV \cite{Bambi1}. Accurate measurements of the reflection spectra of the accretion disk could provide important information about the geometry of the space-time in the strong gravity regime, and thus test the nature of the astrophysical black holes, as well as the possible deviations from the Kerr geometry.

A sample of observational data from seven  Active Galactic Nuclei  observed with
Suzaku  was studied in \cite{Bambi2}, by interpreting the spectrum of the sources with a relativistic reflection component. The results of this analysis are consistent with the hypothesis that the spacetime around these supermassive objects is described by the Kerr geometry. Constraints on the capabilities of X-ray reflection spectroscopy to test the Kerr-nature of astrophysical black holes was considered in \cite{Bambi3}, via the analysis of two NuSTAR observations of Cygnus X-1 in the soft state. It turns out that the final measurement can strongly depend on the assumption of the intensity profile. Moreover, it was concluded that Cygnus X-1 is not a suitable candidate for testing General Relativity using X-ray reflection spectroscopy.  The properties a source with an accretion disk  must have in order to be able to test General Relativity by using X-ray spectroscopy have also been suggested. In principle, supermassive black holes are better candidates than stellar mass black holes. The central object must have fast rotation, with $a_*>0.9$, so that the inner edge of the disk is located closer to the event horizon, and the gravitational effects are stronger. There should be no absorbers between the object and the observer, in order to avoid the astrophysical uncertainties related to the cosmic environment. The data must have a good energy resolution of the iron line, and a broad energy band is necessary to break the parameter degeneracy. The iron line must also be prominent. The accretion luminosity must be between 5\% and 30\% of the Eddington limit, and this condition must hold in order to model the accretion disk as thin. Finally, the corona must have a known geometry, since different coronal geometries are possible. X-ray reflection spectroscopy could provide precision tests of General Relativity in the future once appropriate sources are found, and if precise theoretical models describing the radiation of each component are developed.

The observational evidence for the existence of super massive black holes comes from several astronomical methods. For example, the mass can be accurately determined by analyzing the orbits of stars inside the sphere of the gravitational influence of the black hole \cite{mass}.  An alternative method is represented by the measurement of the diameter of the photon ring encircling the black hole shadow, a method applied for the determination  of the mass of  M87$^{*}$ of the radio galaxy M87 \cite{EHT1,EHT6}.

On the other hand the measurement of the black hole spin is not easy, and it requires the investigation of information coming from around the marginally stable orbits \cite{Abb}. Nonetheless, presently due to the Event Horizon Telescope (EHT), the observational analysis of such a close vicinity to a black hole has become  possible. EHT is a global very long baseline interferometry (VLBI) array observing at 1.3 mm.  EHT observations of M87$^{*}$ have recently provided the first-ever horizon-scale image of a black hole \cite{EHT1,EHT2,EHT3,EHT4,EHT5,EHT6}. These observations show the possibility of the EHT for probing the black hole geometry by timely and spatially resolving the electromagnetic emission coming near the event horizon of black holes. SgrA$^{*}$ has the largest angular size of the gravitational radius and a mass of the order of $M\sim 4\times 10^6M_{\odot}$. It is a black hole candidate with an extremely low luminosity ($L\sim 10^{-9}L_{Edd}$, and with a very low accretion rate $\dot{M}\sim 10^{-8}M_{\odot}\;{\rm yr}^{-1}$.

The above properties and the continuum spectrum of SgrA$^{*}$ can be explained by assuming a radiative inefficient accretion flow \cite{Mar}.  EHT observations for a gas cloud intermittently falling onto a black hole were simulated in \cite{Mori}, where a method for spin measurement based on its relativistic flux variation was proposed.  The light curve of the infalling gas cloud is composed of peaks formed by primary photons that directly reach a distant observer, and by secondary ones reaching the observer after more than one rotation around the black hole.  The black hole spin dependence is detectable in correlated flux densities that are accurately calibrated by baselines between sites with redundant stations.

Hence, one can obtain important astrophysical information from the observation of the motion of the gas streams in
the gravitational field of compact objects. This information does have fundamental theoretical implications, since the study of the accretion and matter flow processes by compact objects is a strong and effective indicator of their physical nature. However, even by taking into account the significant recent advances, up to now the observational results have validated the theoretical predictions of general relativity mostly in a {\it qualitative} way. Despite the present day high precision of the astronomical and astrophysical measurements, still one cannot make a clear observational distinction between the
numerous classes of exotic/compact objects that have been proposed within the
theoretical formalism of general relativity \cite{YuNaRe04}.

Nevertheless, we expect that with the significant improvement of
the already existing imaging observational techniques \cite{EHT1,EHT2,EHT3,EHT4,EHT5,EHT6},  it will also be possible to
obtain definite observational information about the existence of non-trivial black holes or of
naked singularities, and to differentiate these important classes of compact
general relativistic objects.

The black hole solutions of the Einstein equations in vacuum have been extensively investigated.  An important problem concerning black hole solutions is if these spacetimes are nonlinearly stable as solutions of the gravitational field equations.  Another interesting topics is the study of scattering processes on black holes spacetimes. Both these problems can be studied with the help of Teukolsky equation \cite{Teu1,Teu2},  which describe dynamical gravitational, electromagnetic, and neutrino-field perturbations of a rotating black hole. The equations decouple into a single gravitational equation, a single electromagnetic equation, and a single neutrino equation.  The gravitational equation describes the dynamics of the extremal curvature components of the metric in the Newman-Penrose formalism. Around a black hole solution the linearized gravitational equations can be formally decomposed into modes, and this decomposition makes possible to study the so-called mode stability, that is, the existence/non-existence for all metric or curvature components of exponentially growing modes. Up to know most of the researches on mode stability have been performed for the standard solutions of the Einstein vacuum field equations. It would certainly be of interest to also consider the mode stability for the Kerr-Brans-Dicke geometry considered in the present paper.

In the present paper we have convincingly shown that the thermodynamic and
electromagnetic properties (energy flux, temperature
distribution and equilibrium radiation spectrum) of the accretion disks that form around compact objects by gas accretion are different for
naked singularities, Kerr black holes, and non-trivial black holes obtained as rotating solutions of the Brans-Dicke theory for a massless scalar field.  We have obtained a number of observational effects that give some clear
observational signatures that could help to identify observationally and distinguish between different type of compact objects that are the theoretical consequences of the geometric description of gravity. More exactly, by comparing the energy
fluxes emerging from the surface of the gaseous thin accretion disk formed around different types of
black holes and naked singularities having similar masses, we have found that
for some (high) values of the spin parameter and of the model parameter $\gamma $,  the maximal value of the flux is much higher for naked singularities, and the emission region is located more closely to the inner edge of the disk as compared to the Kerr black hole case. In fact, all the physical, geometrical and thermodynamical properties of the disks greatly depend on the values of $\gamma$ and of the spin parameter $a_{\star}$. Similar effects do appear in the behavior of the disk temperature profiles  and of the disk spectra.  Thus, with the future development of the observational techniques
these signatures may provide the possibility of clearly
distinguishing between rotating naked singularities, non-trivial rotating black hole type solutions of the Brans-Dicke theory, and the  Kerr-type black holes of standard general relativity.

\section*{Acknowledgments}

We would like to thank to the anonymous referee for her/his careful reading of the manuscript, and for comments and suggestions that helped us to significantly improve our work. T. H. would like to thank the Yat Sen School of the Sun Yat Sen University in Guangzhou, P. R. China, for the kind hospitality offered during the preparation of this work.

\end{document}